\documentclass[11pt,a4paper]{elsarticle}
\journal{Journal of Computational Physics}
\usepackage{amsmath,amsthm,amsfonts,amssymb,mathrsfs}
\usepackage{graphicx}
\usepackage{subfigure}
\usepackage[latin1]{inputenc}
\usepackage[swedish,english]{babel}

\newcommand{\myvec}[1]{{\boldsymbol#1}}

\begin{document}
\begin{frontmatter}
\title{On a Helmholtz transmission problem in planar domains with
  corners}
\author[label1]{Johan Helsing\corref{cor1}}
\ead{helsing@maths.lth.se}
\address[label1]{Centre for Mathematical Sciences, Lund University,
  Box 118, 221 00 Lund, Sweden}
\author[label2]{Anders Karlsson}
\ead{anders.karlsson@eit.lth.se}
\address[label2]{Electrical and Information Technology, Lund
  University, Box 118, 221 00 Lund, Sweden}
\cortext[cor1]{Corresponding author. URL: http://www.maths.lth.se/na/staff/helsing/}
\begin{abstract}
  A particular mix of integral equations and discretization techniques
  is suggested for the solution of a planar Helmholtz transmission
  problem with relevance to the study of surface plasmon waves. The
  transmission problem describes the scattering of a time-harmonic
  transverse magnetic wave from an infinite dielectric cylinder with
  complex permittivity and sharp edges. Numerical examples illustrate
  that the resulting scheme is capable of obtaining total magnetic and
  electric fields to very high accuracy in the entire computational
  domain.
\end{abstract}
\begin{keyword}
Corner singularity \sep
Helmholtz equation \sep
Nyström discretization \sep 
Scattering \sep
Surface plasmon wave
\MSC[2010] 78M15 \sep 65N38 \sep 35Q60 \sep 35J05 \sep 31A10
\end{keyword}
\end{frontmatter}

\section{Introduction}

This paper is about solving a classic transmission problem for the
Helm\-holtz equation in the plane using integral equation techniques.
A physical interpretation is that an incident time-harmonic transverse
magnetic wave, in a medium with unit permittivity, is scattered from a
homogeneous dielectric cylindrical object with permittivity
$\varepsilon$. The problem is to find the total magnetic field $U$
everywhere.

When $\varepsilon$ is real and positive and when the object boundary
$\Gamma$ is smooth, this problem is uncomplicated. Efficient boundary
integral equations and fast solution techniques have long since been
established and their use in computational physics is standard
practice. See~\cite{Rokh83} for pioneering numerical work
and~\cite{GreeLee12} for an overview of more recent development. The
only issue that, perhaps, still is not completely resolved is how to
compute $U$ and its gradient $\nabla U$ in an appropriate fashion
close to $\Gamma$ in a post-processor~\cite{afKlTorn17,Kloc13}.

When $\varepsilon$ is not real and positive and when $\Gamma$ is not
smooth, the transmission problem gets harder. Issues arise relating to
modeling, the existence and the uniqueness of solutions, and
resolution. A particularly difficult situation is when $\varepsilon$
is real and negative and $\Gamma$ has sharp corners. The excitation of
rapidly oscillating corner fields and their interaction with surface
plasmon waves then make the choice of integral equations and
discretization techniques crucial. To our knowledge, integral equation
methods have not been used in this context, but a finite element
solver has recently been developed~\cite{Annsop16}. This solver, which
relies on so-called perfectly matched layers at the corners, is
capable of producing convergent results also for challenging setups.

We will review integral equations for the Helmholtz transmission
problem and show that a system of equations due to Kleinman and
Martin~\cite{KleiMart88} is well suited for our purposes. In passing
we observe that another, seldom used, integral equation
from~\cite{KleiMart88} is surprisingly efficient when $\varepsilon$ is
real and positive and when the accurate evaluation of $\nabla U$ close
to $\Gamma$ is of concern. The successful use of integral equations in
computations is, of course, coupled to the choice of discretization
scheme. We use standard Nyström discretization, accelerated with
recursively compressed inverse preconditioning, and product
integration for the evaluation of layer potentials close to their
sources~\cite{Hels09,Hels17}. As a result, we can solve the
transmission problem for negative $\varepsilon$ (in a limit sense) in
domains with corners and rapidly obtain corner fields and surface
plasmon waves with a precision of about thirteen digits, even close to
$\Gamma$.

The rest of the paper is organized as follows: Section~\ref{sec:PDE}
presents the transmission problem as a system of partial differential
equations (PDEs). Section~\ref{sec:inteq} reviews some popular
integral equation reformulations which all work well for $\varepsilon$
real and positive. This includes three systems of integral equations
which we call KM0, KM1, and KM2. Section~\ref{sec:disc} is on
discretization. Special emphasis is given to the treatment of
singularities and near-singularities of kernels that occur in field
representations and systems of integral equations. The basic
evaluation strategy is the same as in~\cite{Hels09,HelsHols15}, but
the treatment of the hypersingularity in the gradient of the acoustic
double layer potential operator is new. Section~\ref{sec:unique}
reviews results on the existence and uniqueness of solutions to the
PDEs and to the integral equations of KM0, KM1, and KM2. These issues
are extremely important when $\varepsilon$ is not real and positive
and $\Gamma$ has sharp corners. For inadmissible $\varepsilon$, there
simply is no solution. For a discrete set of other $\varepsilon$, an
inappropriate choice of integral equations may lead to numerical
failure. In Sections~\ref{sec:Drude}, \ref{sec:sing},
and~\ref{sec:cross} we strive to summarize the fascinating physics
which is illustrated by the numerical examples at the end of the
paper.

\section{PDE formulation of the transmission problem}
\label{sec:PDE}

A homogeneous dielectric object, a domain $\Omega_2$ with boundary
$\Gamma$, is embedded in a homogeneous dielectric medium $\Omega_1$ in
the plane $\mathbb{R}^2$. The outward unit normal at position
$r=(x,y)$ on $\Gamma$ is $\nu$. The ratio between the permittivities
in $\Omega_2$ and $\Omega_1$ is $\varepsilon$. An incident plane wave
\begin{gather}
U^{\rm in}(r)=e^{{\rm i}k_1(r\cdot d)}\,,\qquad r\in \mathbb{R}^2\,,
\label{eq:Uin}
\end{gather}
has wavenumber $k_1$, where $\Re{\rm e}\{k_1\}\geq 0$, and direction
$d$. Let the wavenumber in $\Omega_2$ be
\begin{equation}
   k_2=\sqrt{\varepsilon}k_1\,.
\label{eq:k2k1}
\end{equation}
A transmission problem for the Helmholtz equation can now be
formulated: find $U(r)$ which solves the system of PDEs
\begin{align}
\Delta U(r)+k_1^2 U(r)&=0\,,\quad r\in \Omega_1\,,
\label{eq:Hlm1}\\
\Delta U(r)+k_2^2 U(r)&=0\,,\quad r\in \Omega_2\,,
\label{eq:Hlm2}
\end{align}
with boundary conditions
\begin{align}
\lim_{\Omega_1\ni r\to r^\circ} U(r)&=\lim_{\Omega_2\ni r\to r^\circ} U(r)\,,
\quad r^\circ\in\Gamma\,,
\label{eq:Hlm3}\\
\lim_{\Omega_1\ni r\to r^\circ}\varepsilon\nu^\circ\cdot\nabla U(r)&=
\lim_{\Omega_2\ni r\to r^\circ}        \nu^\circ\cdot\nabla U(r)\,,
\quad r^\circ\in\Gamma\,,
\label{eq:Hlm4}\\
  U(r)&=U^{\rm in}(r)+U^{\rm sc}(r)\,,\quad r\in\Omega_1\,,
\label{eq:Hlm5}\\
U^{\rm sc}(r)&=
\dfrac{e^{{\rm i}k\vert r\vert}}{\sqrt{\vert r\vert}}
\left(F(r/\vert r\vert)
+\mathcal{O}\left(\dfrac{1}{\vert r\vert}\right)\right),\, 
\quad\vert r\vert\to\infty\,.
\label{eq:Hlm6}
\end{align}
Here $U^{\rm sc}(r)$ is the scattered field, $F(r/|r|)$ is the
far-field pattern, and~(\ref{eq:Hlm6}) is the two-dimensional analogue
of the radiation condition~\cite[Eq.~(6.22b)]{ColtKres98}.

We are chiefly interested in computing the real fields
\begin{align}
H(r,t)&=\Re{\rm e}\left\{U(r)e^{-{\rm i}t}\right\}\,,
\quad\;\;\; r\in\Omega_1\cup\Omega_2\,,
\label{eq:H}\\
\nabla H(r,t)&=\Re{\rm e}\left\{\nabla U(r)e^{-{\rm i}t}\right\}\,,
\quad r\in\Omega_1\cup\Omega_2\,,
\label{eq:nablaH}
\end{align}
where $t$ denotes time and angular frequency is scaled to one. The
field $H(r,t)$ can be interpreted as a time-harmonic magnetic wave in
a setting where the PDE models a three-dimensional transverse
translation-invariant electromagnetic transmission problem for the
Maxwell equations, with magnetic and electric fields
\begin{align}
{\bf\myvec H}(r)&=U(r)\hat{\myvec z}\,,\\
{\bf\myvec E}(r)&=
\left\{
\begin{array}{ll}
{\rm i}k_1^{-1}\nabla U(r)\times\hat{\myvec z}\,,
&r\in\Omega_1\,,\\
{\rm i}k_1^{-1}\varepsilon^{-1}\nabla U(r)\times\hat{\myvec z}\,,
&r\in\Omega_2\,.
\end{array}
\right.
\label{eq:E}
\end{align}
Here $\hat{\myvec z}$ is a unit vector perpendicular to the plane, the
electric field is scaled with the wave impedance of free space, and
the gradient $\nabla U(r)$ is augmented with a zero third component in
the cross product.

\section{Integral equation formulations}
\label{sec:inteq}

This section reviews some popular integral equation reformulations of
the transmission problem in Section~\ref{sec:PDE} in a uniform
notation. 

\subsection{Bessel functions, kernels, and vectors}
\label{sec:notat}

In what follows, $J_n(x)$ and $Y_n(x)$ are $n$th order Bessel function
of the first and second kind and
\begin{equation}
H_n^{(1)}(x)=J_n(x)+{\rm i}Y_n(x)
\label{eq:Hn1}
\end{equation}
is the $n$th order Hankel function of the first kind. We extend the
definition of the outward unit normal $\nu=\nu(r)$ at a point
$r\in\Gamma$ so that if $r\notin\Gamma$, then $\nu$ is to be
interpreted as an arbitrary unit vector associated with $r$. The
fundamental solution to the Helmholtz equation is taken to be
\begin{equation}
\Phi_k(r,r')=\frac{\rm i}{2}H_0^{(1)}(k|r-r'|)\,,
\label{eq:Phi}
\end{equation}
where $k$ is a wavenumber. We shall also use the double-layer type
kernels with $r'\in\Gamma$ and $\nu'=\nu(r')$ the outward unit normal
at $r'$,
\begin{equation}
D(r,r')=-\frac{\nu\cdot(r-r')}{|r-r'|^2} \qquad {\rm and} \qquad
D(r',r)= \frac{\nu'\cdot(r-r')}{|r-r'|^2}\,.
\label{eq:double}
\end{equation}
The boundary $\Gamma$ has positive orientation and a parameterization
called $r(s)$.

At times we identify vectors $r$, $r'$, $\nu$, $\nu'$ in the real
plane $\mathbb{R}^2$ with points $z$, $\tau$, $n_z$, $n_\tau$ in the
complex plane $\mathbb{C}$. Conjugation of complex quantities is
indicated with an overbar symbol.

\subsection{A standard choice of operators}

We use standard definitions of the single- and double-layer potentials
and their normal derivatives~\cite[Eqs.~(3.8)--(3.11)]{ColtKres98}
\begin{align}
S_k\rho(r)&=\int_{\Gamma}\Phi_k(r,r')\rho(r')\,{\rm d}\ell'\,, 
  \label{eq:Soper}\\
K_k\rho(r)&=
  \int_{\Gamma}\frac{\partial\Phi_k}{\partial\nu'}(r,r')\rho(r')
  \,{\rm d}\ell'\,,\label{eq:Koper}\\
K^{\rm A}_k\rho(r)&=
  \int_{\Gamma}\frac{\partial\Phi_k}{\partial\nu}(r,r')\rho(r')
  \,{\rm d}\ell'\,,\label{eq:KAoper}\\
T_k\rho(r)&=
  \int_{\Gamma}\frac{\partial^2\Phi_k}
  {\partial\nu\partial\nu'}(r,r')\rho(r')\,{\rm d}\ell'\,,
\label{eq:Toper}
\end{align}
where ${\rm d}\ell$ is an element of arc length,
$\partial/\partial\nu=\nu(r)\cdot\nabla$, and
$\partial/\partial\nu'=\nu(r')\cdot\nabla'$. Note
that~\cite{ColtKres98} uses a prefactor ${\rm i}/4$ in the expression
corresponding to~(\ref{eq:Phi}) and a prefactor 2 in the integrals
corresponding to~(\ref{eq:Soper})--(\ref{eq:Toper}). This does not
affect the definitions of $S_k$, $K_k$, $K^{\rm A}_k$, and $T_k$: they
are the same in~(\ref{eq:Soper})--(\ref{eq:Toper}) as
in~\cite[Eqs.~(3.8)--(3.11)]{ColtKres98}.

For $r\in\Gamma$ it holds~\cite[Eqs.~(3.12)--(3.13)]{ColtKres98}
\begin{gather}
K_kK_k-S_kT_k=I\,, 
\label{eq:cald1}\\
K^{\rm A}_kK^{\rm A}_k-T_kS_k=I\,. 
\label{eq:cald2}
\end{gather}

\subsection{The first set of Kleinman--Martin equations}

Kleinman and Martin~\cite[Section~4.1]{KleiMart88} suggest the field
representation
\begin{gather}
U(r)=U^{\rm in}(r)+\frac{1}{2}K_{k_1}\mu(r)+\frac{1}{2}S_{k_1}\rho(r)\,,
\quad r\in\Omega_1\,,
\label{eq:U1KM1}\\
U(r)=\frac{\varepsilon}{2}K_{k_2}\mu(r)+\frac{c}{2}S_{k_2}\rho(r)\,,
\quad r\in\Omega_2\,,
\label{eq:U2KM1}
\end{gather}
where $c$ is a constant such that $c+\varepsilon\ne 0$ and
\begin{equation}
\arg(c)=\left\{
\begin{array}{lll}
\arg(\varepsilon k_2)    &\mbox{if}&\Re{\rm e}\{k_1\}\ge 0\,,\\
\arg(\varepsilon k_2)-\pi&\mbox{if}&\Re{\rm e}\{k_1\}< 0\,.
\end{array}
\right.
\label{eq:cchoice0}
\end{equation}
The corresponding system of integral equations is
\begin{equation}
\begin{bmatrix}
I-\alpha_2K_{k_2}+\alpha_1K_{k_1} & -\alpha_1(cS_{k_2}-S_{k_1}) \\
\alpha_4(T_{k_2}-T_{k_1}) &I+c\alpha_3K^{\rm A}_{k_2}-\alpha_4K^{\rm A}_{k_1} 
\end{bmatrix}
\begin{bmatrix}
\mu(r)\\
\rho(r)
\end{bmatrix}
=
\begin{bmatrix}
 f_1(r)\\
 f_2(r)
\end{bmatrix}\,,
\label{eq:KM1sys}
\end{equation}
with $r\in\Gamma$ and
\begin{gather}
f_1(r)=-2\alpha_1U^{\rm in}(r)\,, \qquad
f_2(r)= 2\alpha_4\frac{\partial U^{\rm in}}{\partial\nu}(r)\,, 
\label{eq:f1f2c}\\
\alpha_1=\frac{1}{1+\varepsilon}\,,\quad
\alpha_2=\frac{\varepsilon}{1+\varepsilon}\,,\quad
\alpha_3=\frac{1}{c+\varepsilon}\,,\quad
\alpha_4=\frac{\varepsilon}{c+\varepsilon}\,.
\label{eq:alphasc}
\end{gather}
The choice $c$ in~(\ref{eq:cchoice0}) guarantees the uniqueness of the
solution $\mu$, $\rho$ to~(\ref{eq:KM1sys}) under certain conditions.
See, further, Section~\ref{sec:KM1unique}.

The equations~(\ref{eq:U1KM1}), (\ref{eq:U2KM1}),
and~(\ref{eq:KM1sys}) will be referred to as the KM1 representation
and system, or simply the KM1 equations.

\subsection{The second set of Kleinman--Martin equations}

Kleinman and Martin~\cite[Section~4.2]{KleiMart88} also suggest the field
representation
\begin{gather}
U(r)=U^{\rm in}(r)+\frac{1}{2}K_{k_1}\mu(r)-\frac{1}{2}S_{k_1}\rho(r)\,,
\quad r\in\Omega_1\,,
\label{eq:U1KM2}\\
U(r)=-\frac{1}{2}K_{k_2}\mu(r)+\frac{\varepsilon}{2}S_{k_2}\rho(r)\,,
\quad r\in\Omega_2\,,
\label{eq:U2KM2}
\end{gather}
where the layer densities $\mu$ and $\rho$ have the physical
interpretations
\begin{gather}
\mu(r) =\lim_{\Omega_1\cup\Omega_2\ni r\to r^\circ} U(r)\,,
\quad r^\circ\in\Gamma\,,\\
\rho(r)=\lim_{\Omega_1\ni r\to r^\circ}\nu^\circ\cdot\nabla U(r)\,,
\quad r^\circ\in\Gamma\,.
\end{gather}

Taking the limit $r\to\Gamma$ in~(\ref{eq:U1KM2}) and~(\ref{eq:U2KM2})
gives two integral equations. The limits of the gradients
in~(\ref{eq:U1KM2}) and~(\ref{eq:U2KM2}) give two additional
equations. In~\cite[Section~4.2]{KleiMart88}, these four equations are
combined into the system of integral equations
\begin{equation}
\begin{bmatrix}
I+\alpha_1K_{k_2}-\alpha_2 K_{k_1} & -\alpha_2(S_{k_2}-S_{k_1})\\
\alpha_1(T_{k_2}-T_{k_1}) &I-\alpha_2K^{\rm A}_{k_2}+\alpha_1K^{\rm A}_{k_1}
\end{bmatrix}
\begin{bmatrix}
\mu(r)\\
\rho(r)
\end{bmatrix}
=
\begin{bmatrix}
 f_1(r)\\
 f_2(r)
\end{bmatrix}\,,
\label{eq:KM2sys}
\end{equation}
with $r\in\Gamma$ and
\begin{gather}
f_1(r)=2\alpha_2U^{\rm in}(r)\,, \qquad
f_2(r)=2\alpha_1\frac{\partial U^{\rm in}}{\partial\nu}(r)\,,\\
\alpha_1=\frac{1}{1+\varepsilon}\,,\qquad
\alpha_2=\frac{\varepsilon}{1+\varepsilon}\,.
\end{gather}
The equations~(\ref{eq:U1KM2}), (\ref{eq:U2KM2}),
and~(\ref{eq:KM2sys}) will be referred to as the KM2 representation
and system, or simply the KM2 equations.

An advantage with the KM2 representation is that, by adding the null
fields of~\cite[Eqs.~(3.15a) and~(3.14b)]{KleiMart88}
\begin{gather}
0=-\frac{1}{2}K_{k_2}\mu(r)+\frac{\varepsilon}{2}S_{k_2}\rho(r)\,,
\quad r\in\Omega_1\,,
\label{eq:U2KM2Null}\\
0=U^{\rm in}(r)+\frac{1}{2}K_{k_1}\mu(r)-\frac{1}{2}S_{k_1}\rho(r)\,,
\quad r\in\Omega_2\,,
\label{eq:U1KM2Null}
\end{gather}
to~(\ref{eq:U1KM2}) and~(\ref{eq:U2KM2}), the representation of $U(r)$
can be written
\begin{equation}
U(r)=U^{\rm in}(r)
  -\frac{1}{2}\left(K_{k_2}-K_{k_1}\right)\mu(r)
  +\frac{1}{2}\left(\varepsilon S_{k_2}-S_{k_1}\right)\rho(r)\,,
\quad r\in\Omega_1\cup\Omega_2\,.
\label{eq:U3KM2}
\end{equation}
The representation~(\ref{eq:U3KM2}) contains the difference operator
$K_{k_2}-K_{k_1}$ whose kernel is smoother close to $\Gamma$ than
those of the individual operators $K_{k_2}$ and $K_{k_1}$
in~(\ref{eq:U1KM1}), (\ref{eq:U2KM1}), (\ref{eq:U1KM2}),
and~(\ref{eq:U2KM2}). As we shall see in Section~\ref{sec:fgf}, the
stabilizing effect of using~(\ref{eq:U3KM2}) is particularly
pronounced when the gradient field $\nabla U(r)$ is computed.

\subsection{The equations used by Greengard and Lee}

Greengard and Lee use the field
representation~\cite[Eq.~(8)]{GreeLee12} that results from setting
$c=1$ in the KM1 equations. The corresponding system of integral
equations~\cite[Eq.~(11)]{GreeLee12} is therefore identical
to~(\ref{eq:KM1sys}) with $c=1$. The equations~(\ref{eq:U1KM1}),
(\ref{eq:U2KM1}), and~(\ref{eq:KM1sys}) with $c=1$ will be referred to
as the KM0 representation and system, or simply the KM0 equations.

\subsection{The Kress--Roach and Müller--Rokhlin equations}

Kress and Roach~\cite{KresRoac78} and Rokhlin~\cite{Rokh83} study
the transmission problem of Section~\ref{sec:PDE} with the boundary
conditions~(\ref{eq:Hlm3}) and~(\ref{eq:Hlm4}) replaced by
\begin{gather}
  \lim_{\Omega_1\ni r\to r^\circ}U(r)=\lim_{\Omega_2\ni r\to
    r^\circ}\varepsilon U(r)\,,
  \quad r^\circ\in\Gamma\,,\\
  \lim_{\Omega_1\ni r\to r^\circ}\nu^\circ\cdot\nabla U(r)=
  \lim_{\Omega_2\ni r\to r^\circ}\nu^\circ\cdot\nabla U(r)\,, \quad
  r^\circ\in\Gamma\,.
\end{gather}

The field representation and system in~\cite{KresRoac78} is similar to
the KM1 equations, but contains two free parameters $c_1$ and $c_2$.
One can view the KM1 equations as a simplification of the Kress--Roach
equations and we will not investigate the latter equations numerically
in this work.

The field representation for $U(r)$ in~\cite{Rokh83} is
\begin{gather}
U(r)=U^{\rm in}(r)+\frac{1}{2}K_{k_1}\mu(r)+\frac{1}{2}S_{k_1}\rho(r)\,,
\quad r\in\Omega_1\,,
\label{eq:U1MR}\\
U(r)=\frac{1}{2}K_{k_2}\mu(r)+\frac{1}{2\varepsilon}S_{k_2}\rho(r)\,,
\quad r\in\Omega_2\,.
\label{eq:U2MR}
\end{gather}
This representation gives rise to a system of integral equations,
often called the Müller--Rokhlin equations, which is identical to the
KM1 system with $c=1$ and the KM0 system. Since the representation for
$U(r)$ in~(\ref{eq:U2MR}) differs from $U(r)$ in~(\ref{eq:U2KM1}) with
$c=1$, the expression~(\ref{eq:H}) for $H(r,t)$ changes into
\begin{equation}
H(r,t)=\left\{
\begin{array}{ll}
\Re{\rm e}\left\{U(r)e^{-{\rm i}t}\right\}\,,& r\in\Omega_1\,,\\
\Re{\rm e}\left\{\varepsilon U(r)e^{-{\rm i}t}\right\}\,,& r\in\Omega_2\,,
\end{array}
\right.
\end{equation}
and the expression~(\ref{eq:nablaH}) for $\nabla H(r,t)$ undergoes an
analogous change.

\section{Discretization}
\label{sec:disc}

We discretize and solve the integral equations of
Section~\ref{sec:inteq} using Nyström discretization with composite
$n_{\rm pt}$-point Gauss--Legendre quadrature as underlying
quadrature. Most often we choose $n_{\rm pt}=16$.

When the boundary $\Gamma$ contains corners, the Nyström scheme is
accelerated and stabilized with recursively compressed inverse
preconditioning (RCIP). The RCIP technique accomplishes, in linear or
sublinear time, a lossless compression of Fredholm second kind
integral equations discretized on meshes increasingly {\it refined} in
the direction toward corner vertices. The final preconditioned system
is solved for transformed layer densities, represented by their values
at discretization points only on a {\it coarse} mesh on $\Gamma$. See
the compendium~\cite{Hels17} for a thorough review of RCIP
acceleration of Nyström schemes. See~\cite{HelsKarl16,HelsPerf13} for
applications of these techniques to the solution of integral equations
that are similar to those of Section~\ref{sec:inteq}.
See~\cite[Section 6.2-6.3]{Hels11} for details on performance
enhancement involving Newton's method and homotopy, and for a
discussion of the correspondence between traditional mesh refinement
and the number of recursion steps used in advanced implementations of
RCIP. See~\cite[Section 7.2.2]{HelsPerf18} for general comments on how
the need for local resolution, $n_{\rm pt}$, depends on $\varepsilon$.

When operator kernels $G(r,r')$ contain singularities, or
near-singularities, we replace the Gauss--Legendre quadrature, on
quadrature panels affected, with a product integration scheme. This
scheme was first described in~\cite[Section~2]{Hels09} and further
developed in~\cite[Section~6]{HelsHols15} and has as its key the
construction of a split
\begin{multline}
G(r,r')\,{\rm d}\ell'=G_0(r,r')\,{\rm d}\ell'
+\log|r-r'|G_{\rm L}(r,r')\,{\rm d}\ell'\\
+\Re{\rm e}\left\{\frac{G_{\rm C}(z,\tau)\,{\rm d}\tau}
                 {{\rm i}(\tau-z)}\right\}
+\Re{\rm e}\left\{\frac{G_{\rm H}(z,\tau)\,{\rm d}\tau}
                 {{\rm i}(\tau-z)^2}\right\}\,,
\label{eq:ex3}
\end{multline}
where $G_0(r,r')$, $G_{\rm L}(r,r')$, $G_{\rm C}(z,\tau)$, and $G_{\rm
  H}(z,\tau)$ are smooth functions and complex notation is used as
explained in Section~\ref{sec:notat}. Product integration weights for
the kernels of~(\ref{eq:ex3}) are then obtained using analytical
methods and recursion or, when $r\in\Gamma$, alternatively by local
regularization~\cite[Section~2]{Hels09}. The scheme requires explicit
formulas for $G(r,r')$, $G_{\rm L}(r,r')$, $G_{\rm C}(z,\tau)$, and
$G_{\rm H}(z,\tau)$, while $G_0(r,r')$ needs only to be known if
$r\in\Gamma$ and then only in the limit $r'\to r$.

In the remainder of this section the singular nature of the kernels of
$S_k$, $K_k$, $K^{\rm A}_k$, and $T_k$ is explored, so that splits of
the form~(\ref{eq:ex3}) can be constructed.

\subsection{Expansions of $Y_n(x)$ around $x=0$}

The following series expansions~\cite{bessec} of $Y_n(x)$ around $x=0$
are useful:
\begin{equation}
Y_0(x)=\frac{2}{\pi}J_0(x)\log\left(\frac{x}{2}\right)
-\frac{2\psi(1)}{\pi}
-\frac{1}{\pi}\sum_{j=1}^\infty\frac{(-1)^j2\psi(j+1)}
{j!j!}\left(\frac{x}{2}\right)^{2j}\,,
\label{eq:Y0}
\end{equation}
\begin{multline}
Y_1(x)=\frac{2}{\pi}J_1(x)\log\left(\frac{x}{2}\right)
-\frac{2}{\pi x}\\
-\frac{1}{\pi}\sum_{j=0}^\infty\frac{(-1)^j(\psi(j+1)+\psi(j+2))}
{j!(j+1)!}\left(\frac{x}{2}\right)^{2j+1}\,,
\label{eq:Y1}
\end{multline}
\begin{multline}
Y_2(x)=\frac{2}{\pi}J_2(x)\log\left(\frac{x}{2}\right)
-\frac{4}{\pi x^2}-\frac{1}{\pi}\\
-\frac{1}{\pi}\sum_{j=0}^\infty\frac{(-1)^j(\psi(j+1)+\psi(j+3))}
{j!(j+2)!}\left(\frac{x}{2}\right)^{2j+2}\,,
\label{eq:Y2}
\end{multline}
where $\psi(\cdot)$ is the digamma function.

\subsection{The kernel of $S_k$}

For arbitrary wavenumber $k$ and with $r$ not necessarily on $\Gamma$
we have
\begin{equation}
S_k(r,r')=\frac{\rm i}{2}H_0^{(1)}(k|r-r'|)\,.
\end{equation}
Using~(\ref{eq:Hn1}) and~(\ref{eq:Y0}) one can split
$G(r,r')=S_k(r,r')$ in the form~(\ref{eq:ex3}) with
\begin{equation}
G_{\rm L}(r,r')=-\frac{1}{\pi}J_0(k|r-r'|)\,,
\end{equation}
$G_{\rm C}=0$, $G_{\rm H}=0$, and general limit
\begin{equation}
\lim_{r'\to r}G_0(r,r')=\frac{\rm i}{2}
-\frac{1}{\pi}\left(\log(k/2)-\psi(1)\right)\,.
\end{equation}

\subsection{The kernel of $K_k$}

For arbitrary wavenumber $k$ and with $r$ not necessarily on $\Gamma$
we have
\begin{equation}
K_k(r,r')=\frac{\rm i}{2}k|r-r'|H_1^{(1)}(k|r-r'|)D(r',r)\,,
\end{equation}
with $D(r',r)$ as in~(\ref{eq:double}). In the limit of $k\to 0$ this
means
\begin{equation}
K_0\rho(r)=\frac{1}{\pi}\int_\Gamma D(r',r)\rho(r')\,{\rm d}\ell'
          =-\Re{\rm e}\left\{\frac{1}{\pi{\rm i}}
           \int_\Gamma\frac{\rho(\tau)\,{\rm d}\tau}{\tau-z}\right\}\,,
\label{eq:K0}
\end{equation}
which is the Neumann--Poincaré operator with negative sign.
Using~(\ref{eq:Hn1}) and~(\ref{eq:Y1}) one can split
$G(r,r')=K_k(r,r')$ in the form~(\ref{eq:ex3}) with
\begin{align}
G_{\rm L}(r,r')  &=-\frac{1}{\pi}k|r-r'|J_1(k|r-r'|)D(r',r)\,,\\
G_{\rm C}(z,\tau)&=-\frac{1}{\pi}\,,
\label{eq:KC}
\end{align}
$G_{\rm H}=0$, and general limit
\begin{equation}
\lim_{r'\to r}G_0(r,r')=0\,.\\
\end{equation}

For $r$ on smooth $\Gamma$ and with $G_{\rm C}$ as in~(\ref{eq:KC}),
the third term on the right hand side of~(\ref{eq:ex3}) is smooth and
should be included in the first term. Then $G_{\rm C}=0$ and
\begin{equation}
\lim_{r'\to r}G_0(r,r')=\frac{(\nu\cdot\ddot{r})}{2\pi|\dot{r}|^2}\,,
\end{equation}
where $\dot{r}={\rm d}r(s)/{\rm d}s$ and $\ddot{r}={\rm d}^2r(s)/{\rm
  d}s^2$. 

Note that, thanks to $k$-independence in~(\ref{eq:KC}), the term
associated with $G_{\rm C}$ in~(\ref{eq:ex3}) cancels out in
difference operators $K_{k_2}-K_{k_1}$.

\subsection{The kernel of $K^{\rm A}_k$}

For arbitrary wavenumber $k$ and with $r$ not necessarily on $\Gamma$
we have
\begin{equation}
K^{\rm A}_k(r,r')=\frac{\rm i}{2}k|r-r'|H_1^{(1)}(k|r-r'|)D(r,r')\,,
\end{equation}
with $D(r,r')$ as in~(\ref{eq:double}). In the limit of $k\to 0$ this
means
\begin{equation}
K^{\rm A}_0\rho(r)=\frac{1}{\pi}\int_\Gamma D(r,r')\rho(r')\,{\rm d}\ell'
         =\Re{\rm e}\left\{\frac{1}{\pi{\rm i}}
   \int_\Gamma\frac{n_z\bar{n}_\tau\rho(\tau)\,{\rm d}\tau}{\tau-z}\right\}\,.
\label{eq:KA0}
\end{equation}

Using~(\ref{eq:Hn1}) and~(\ref{eq:Y1}) one can split $G(r,r')=K^{\rm
  A}_k(r,r')$ in the form~(\ref{eq:ex3}) with
\begin{align}
G_{\rm L}(r,r')  &=-\frac{1}{\pi}k|r-r'|J_1(k|r-r'|)D(r,r')\,,\\
G_{\rm C}(z,\tau)&=\frac{n_z\bar{n}_\tau}{\pi}\,,
\label{eq:KAC}
\end{align}
$G_{\rm H}=0$, and general limit
\begin{equation}
\lim_{r'\to r}G_0(r,r')=0\,.
\end{equation}

For $r$ on smooth $\Gamma$ and with $G_{\rm C}$ as in~(\ref{eq:KAC}),
the third term on the right hand side of~(\ref{eq:ex3}) is smooth and
should be included in the first term. Then $G_{\rm C}=0$ and
\begin{equation}
\lim_{r'\to r}G_0(r,r')=\frac{(\nu\cdot\ddot{r})}{2\pi|\dot{r}|^2}\,.
\end{equation}

\subsection{The kernel of $T_k$}

For arbitrary wavenumber $k$ and with $r$ not necessarily on $\Gamma$
we have
\begin{align}
T_k(r,r')&=\frac{\rm i}{2}k|r-r'|H_1^{(1)}(k|r-r'|)
\frac{(\nu\cdot\nu')}{|r-r'|^2}\nonumber\\
    &+\frac{\rm i}{2}(k|r-r'|)^2H_2^{(1)}(k|r-r'|)D(r,r')D(r',r)\,.
\end{align}
Using~(\ref{eq:Hn1}) and~(\ref{eq:Y2}) one can split
$G(r,r')=T_k(r,r')$ in the form~(\ref{eq:ex3}) with
\begin{align}
G_{\rm L}(r,r')&=-\frac{k}{\pi}J_1(k|r-r'|)\frac{(\nu\cdot\nu')}{|r-r'|}
\nonumber\\
            &\quad\, -\frac{1}{\pi}(k|r-r'|)^2J_2(k|r-r'|)D(r,r')D(r',r)\,,\\
G_{\rm C}(z,\tau)&=-\frac{k^2}{2\pi}
\Re{\rm e}\left\{n_z(\bar{\tau}-\bar{z})\right\}\,,
\label{eq:TC}\\
G_{\rm H}(z,\tau)&=-\frac{n_z}{\pi}\,.
\label{eq:TH}
\end{align}
and general limit
\begin{equation}
\lim_{r'\to r}G_0(r,r')=\frac{\rm i}{4}k^2
-\frac{1}{4\pi}k^2(2\log(k/2)-2\psi(1)-1)\,.
\label{eq:G_0T}
\end{equation}

For $r$ on smooth $\Gamma$ and with $G_{\rm C}$ as in~(\ref{eq:TC}),
the third term on the right hand side of~(\ref{eq:ex3}) is smooth, has
zero limit as $r'\to r$, and should be included in the first term.
Then $G_{\rm C}=0$, but~(\ref{eq:G_0T}) is unaffected.

Note that, thanks to $k$-independence in~(\ref{eq:TH}), the term
associated with $G_{\rm H}$ in~(\ref{eq:ex3}) cancels out in
difference operators $T_{k_2}-T_{k_1}$.

\section{Existence and uniqueness of solutions}
\label{sec:unique}

This section collects known results on the existence and the
uniqueness of solutions to the PDE~(\ref{eq:Hlm1})--(\ref{eq:Hlm6})
and to the system of integral equations called KM0, KM1, and KM2 in
Section~\ref{sec:inteq}.

\subsection{Uniqueness: the PDE}

According to the uniqueness theorem of~\cite[Section~2]{KleiMart88},
if $\Gamma$ is smooth, if $k_1\ne 0$, and if~(\ref{eq:k2k1}) is
assumed, a solution to~(\ref{eq:Hlm1})--(\ref{eq:Hlm6}) is unique if
\begin{equation}
 0\le\arg(k_1)<\pi\,,
\quad |\varepsilon|\ne\infty\,,\quad 0\le\arg(\varepsilon k_1)\le\pi\,.
\label{eq:Klein}
\end{equation}
According to~\cite[Theorem~3.1]{KresRoac78}, if $\Gamma$ is smooth, if
$k_1,k_2\ne 0$, and if~(\ref{eq:k2k1}) is assumed, a solution
to~(\ref{eq:Hlm1})--(\ref{eq:Hlm6}) is unique if
\begin{equation}
0\le\arg(k_1),\arg(k_2)\le\pi/2\quad \mbox{or}\quad
\pi/2\le\arg(k_1),\arg(k_2)<\pi\,.
\label{eq:Kress}
\end{equation}
The two sets of conditions~(\ref{eq:Klein}) and~(\ref{eq:Kress})
overlap, but are not identical. For example, if $\arg(k_1)=\pi/2$ and
$0<\arg(k_2)<\pi/4$ then~(\ref{eq:Kress}) holds but
not~(\ref{eq:Klein}) -- a fact we think is due to a flaw in the proof
of~\cite[Theorem~3.1]{KresRoac78} which affects the analysis of the
solvability of~\cite[Eq.~(4.5)]{KresRoac78}. In the present work we
are chiefly interested in $\arg(k_1)=0$, $0\le\arg(k_2)\le\pi/2$. Then
both~(\ref{eq:Klein}) and~(\ref{eq:Kress}) guarantee that a solution
to~(\ref{eq:Hlm1})--(\ref{eq:Hlm6}) is unique if $\Gamma$ is smooth.

\subsection{Unique solvability: the systems of integral equations}
\label{sec:KM1unique}

According to~\cite[Theorem~4.1]{KleiMart88}, if $\Gamma$ is smooth,
if~(\ref{eq:Klein}) holds, and if $c$ is as in~(\ref{eq:cchoice0}),
then the KM1 system~(\ref{eq:KM1sys}) is uniquely solvable. The unique
solution $\mu$, $\rho$ gives, via~(\ref{eq:U1KM1})
and~(\ref{eq:U2KM1}), a unique solution
to~(\ref{eq:Hlm1})--(\ref{eq:Hlm6}). As a consequence, the KM0 system
and the Müller-Rokhlin equations, which correspond to the KM1 system
with $c=1$, are also uniquely solvable if both $k_1$ and $k_2$ are
real and positive. In our numerical experiments with KM1 in
Section~\ref{sec:numex}, where $\Re{\rm e}\{k_1\}\ge 0$, we choose $c$
in accordance with~(\ref{eq:cchoice0}) as
\begin{equation}
c=\varepsilon k_2/\lvert\varepsilon k_2\rvert\,.
\label{eq:cchoice}
\end{equation}

According to~\cite[Theorems~4.2 and~4.3]{KleiMart88}, if $\Gamma$ is
smooth and if both $k_1$ and $k_2$ are real and positive, then the KM2
system~(\ref{eq:KM2sys}) is uniquely solvable. The solution
gives, via~(\ref{eq:U1KM2}) and~(\ref{eq:U2KM2}), a unique solution
to~(\ref{eq:Hlm1})--(\ref{eq:Hlm6}).

\subsection{True and false eigenwavenumbers}
\label{sec:trufaleig}

When $k_1$ and $\varepsilon$ are such that the
conditions~(\ref{eq:Klein}) are violated, a solution
to~(\ref{eq:Hlm1})--(\ref{eq:Hlm6}) may, or may not, be unique. The
same applies to solutions to the systems of integral equations of
Section~\ref{sec:inteq} if the conditions on unique solvability of
Section~\ref{sec:KM1unique} are violated.

Assume now that the conditions on unique solvability of
Section~\ref{sec:KM1unique} are violated for a given system of
integral equations in Section~\ref{sec:inteq}. If, for some $k_1$ and
$\varepsilon$, we numerically detect a non-trivial homogeneous
solution to that system, we call $k_1$ an {\it eigenwavenumber}.
Eigenwavenumbers can be of two types: those that correspond to
non-vanishing eigenfields $U(r)$ that satisfy the boundary
conditions~(\ref{eq:Hlm3}) and~(\ref{eq:Hlm4}) and those that
correspond to $U(r)=0$ or violate the boundary
condition~(\ref{eq:Hlm3}). We call the former type {\it true
  eigenwavenumbers} and the latter type {\it false eigenwavenumbers}.
False eigenwavenumbers that correspond to eigenfields that
violate~(\ref{eq:Hlm3}) can only occur for the KM2 equations since
$U(r)$ in the KM0 and KM1 equations, by construction, always
satisfies~(\ref{eq:Hlm3}).

\subsection{Existence: the KM1 system on a boundary with corners}

This section discusses some issues related to the existence of
solutions to the KM1 system~(\ref{eq:KM1sys}) when $\Gamma$ has
corners, $\varepsilon$ is close to or on the negative real axis, and $k_1$ is
real and positive so that $c\approx-{\rm i}$ according
to~(\ref{eq:cchoice}).

In view of the singular nature of the kernels of the integral
operators in Section~\ref{sec:disc}, and when $r\in\Gamma$, the KM1
system can be written in the form
\begin{equation}
\begin{bmatrix}
I+\lambda_1 K_0+C_1 & C_2  \\
 C_3              & I+\lambda_2 K^{\rm A}_0 +C_4
\end{bmatrix}
\begin{bmatrix}
\mu(r)\\
\rho(r)
\end{bmatrix}
=
\begin{bmatrix}
 f_1(r)\\
 f_2(r)
\end{bmatrix}\,,
\label{eq:KM1sysc}
\end{equation}
where $C_1$, $C_2$, $C_3$, and $C_4$ are compact integral operators on
$\Gamma$ also in the presence of corners, $K_0$ and $K^{\rm A}_0$ are
as in~(\ref{eq:K0}) and~(\ref{eq:KA0}), and
\begin{equation}
\lambda_1=\frac{1-\varepsilon}{1+\varepsilon}\,,\qquad
\lambda_2=\frac{c-\varepsilon}{c+\varepsilon}\,.
\label{eq:lambda12}
\end{equation}
The operators $K_0$ and $K^{\rm A}_0$ are singular bounded integral
operators on $\Gamma$. The operator $K^{\rm A}_0$ is the same as the
operator denoted $K$ in~\cite[Eq.~(10)]{HelsPerf13}.

The system~(\ref{eq:KM1sysc}) is a compact perturbation of the
de-coupled system
\begin{equation}
\begin{bmatrix}
I+\lambda_1K_0 & 0  \\
 0            & I+\lambda_2 K^{\rm A}_0
\end{bmatrix}
\begin{bmatrix}
\mu(r)\\
\rho(r)
\end{bmatrix}
=
\begin{bmatrix}
 f_1(r)\\
 f_2(r)
\end{bmatrix}\,,
\label{eq:KM1sysdc}
\end{equation}
whose spectral properties, including the essential spectrum of $K_0$
and $K^{\rm A}_0$ in planar domains with corners, have been analyzed
in~\cite{PerfPuti17}.

In particular, when $\Gamma$ has a corner with opening angle $\theta$
and when $\varepsilon$ of~(\ref{eq:lambda12}) is real and such that
\begin{equation}
-\left\lvert 1-\frac{\theta}{\pi}\right\rvert < -\frac{1}{\lambda_1} <
 \left\lvert 1-\frac{\theta}{\pi}\right\rvert
\quad
\Leftrightarrow
\quad
 \frac{\lvert\pi-\theta\rvert+\pi}{\lvert\pi-\theta\rvert-\pi}
<\varepsilon<
 \frac{\lvert\pi-\theta\rvert-\pi}{\lvert\pi-\theta\rvert+\pi}
\,,
\label{eq:z1cont}
\end{equation}
then~(\ref{eq:KM1sysdc}) does not in general have a solution $\mu$ in
the fractional Sobolev space $H^{1/2}(\Gamma)$, see~\cite{Ding96},
corresponding to boundary values of potentials with finite absolute
energy in the sense of~\cite[Section~4]{HelsPerf13}. There exists,
however, a solution $\mu\in H^{1/2}(\Gamma)$ for $\lambda_1$
arbitrarily close to, but not on, the real axis. We remark that
similar restrictions on the solvability of~(\ref{eq:KM1sysdc}) apply
to $\rho$ if $-1/\lambda_2$ is real and in the interval specified
by~(\ref{eq:z1cont}). Since, by assumption on $\varepsilon$ and $k_1$,
the parameter $\lambda_2$ of~(\ref{eq:lambda12}) is not close to the
real axis and we disregard this possibility.

The essential spectrum of an integral operator is invariant under
compact perturbations. Therefore the solvability analysis
for~(\ref{eq:KM1sysdc}) also applies to~(\ref{eq:KM1sysc}).
When~(\ref{eq:z1cont}) holds in our numerical examples of
Section~\ref{sec:numex}, we then add a small imaginary number ${\rm
  i}\delta$, $\delta>0$, to $\varepsilon$ and solve~(\ref{eq:KM1sysc})
in the limit $\delta\to 0^+$. This corresponds to $\lambda_1$
approaching the real axis from below in the complex plane. The limit
solutions are indicated with a plus-sign superscript.

\section{The Drude model and surface plasmon waves}
\label{sec:Drude}

We are interested in wave phenomena which under certain conditions
appear in metallic objects with sharp edges. Let us assume that there
is vacuum in $\Omega_1$, that $k_1$ is real and positive, and that
$\Omega_2$ is a metal. Certain metals, such as silver, have
permittivities that often are well approximated by the Drude model,
see~\cite[Eq.~(7.58)]{Jackson1999}, which in our context reads
\begin{equation}
\varepsilon=1-\frac{k_{\rm p}^2}{k_1^2+{\rm i}k_1\gamma}\,.
\label{eq:Drude}
\end{equation}
Here $k_{\rm p}$ is the plasma wavenumber and $\gamma\ge 0$ is a
damping constant. The equation says that $\varepsilon$ becomes real
and negative when $k_1<k_{\rm p}$ and $\gamma\to 0^+$. We mention the
Drude model merely to explain why real and negative $\varepsilon$ can
occur for real and positive $k_1$.

When $\varepsilon$ is real the time average of the electric and the
magnetic energy densities, normed by the vacuum permeability, are
\begin{align}
\langle w_{\rm el}(r)\rangle&=
     \frac{1}{4 k_i^2}\vert\nabla U(r)\vert^2\,,
\quad r\in\Omega_i\,,\quad i=1,2\,,\\
\langle w_{\rm ma}(r)\rangle&=
     \frac{1}{4}\vert U(r)\vert^2\,,\quad r\in\mathbb{R}^2\,.
\end{align}
This means that if $\varepsilon<0$, then $\langle w_{\rm
  el}(r)\rangle$ is negative inside $\Omega_2$ while $\langle w_{\rm
  ma}(r)\rangle$ is positive and electromagnetic waves cannot
propagate in $\Omega_2$. It is, however, possible for so-called {\it
  surface plasmon waves} to propagate in a direction along $\Gamma$.
This is illustrated in the numerical examples of
Section~\ref{sec:spw}, below, where surface plasmon waves are excited
and propagate along $\Gamma$ of a bounded object. The properties of
these waves resemble those of surface plasmon waves along planar
surfaces, and this resemblance is enhanced as the wavelength of the
surface plasmon waves decreases.

The main results for surface plasmon waves on planar surfaces,
pertinent also for curved $\Gamma$, are as follows, see~\cite[Appendix
I]{Raet88}: The surface plasmon waves can occur only for
$\varepsilon<-1$, they are evanescent (which roughly means
exponentially decaying) in directions perpendicular to $\Gamma$, and
propagate along $\Gamma$ with wavenumber
\begin{equation}
k_{\rm sp}=k_1\sqrt{\frac{|\varepsilon|}{|\varepsilon|-1}}\,,
\quad\varepsilon<-1\,.
\label{eq:spl}
\end{equation}

Surface plasmon waves can always propagate when $\varepsilon<-1$ and
$k_1>0$, but their excitation by an incident wave $U^{\rm in}(r)$
requires special couplers or that $\Gamma$ is somehow
rough~\cite[Sections~2.2 and 6.7]{Raet88}. Surface plasmon waves are
particularly easy to excite when $\Gamma$ has corners and
$\varepsilon$ is infinitely close to the interval
where~(\ref{eq:z1cont}) holds. That is, when $\varepsilon$ approaches
the interval
\begin{equation}
\frac{\lvert\pi-\theta\rvert+\pi}{\lvert\pi-\theta\rvert-\pi}
 < \varepsilon < -1
\label{eq:excite}
\end{equation}
from above in the complex plane. Note that, in~(\ref{eq:spl}), the
wavenumber $k_{\rm sp}$ diverges as $\varepsilon\to -1^-$. More
details on surface plasmon waves can, for example, be found
in~\cite{BaDeEb03,WangShen06}. Their excitation by corners is the
topic of the next section.

\section{Singular fields at corners}
\label{sec:sing}

The presence of corners on $\Gamma$ has a great influence on the
excitation of surface plasmon waves and, as a consequence, on
scattering and absorption cross sections. The mechanism behind this is
governed by certain singular eigenfields which can be determined by
quasi-static analysis, see~\cite{Annsop16}. We now briefly review some
results on this topic.

\subsection{Quasi-static eigenfields}
\label{sec:quasi}

In the limit $k_1\to 0$ and for certain $\varepsilon$, the
transmission problem of Section~\ref{sec:PDE} can allow for magnetic
eigenfields. These eigenfields are non-trivial solutions to
\begin{align}
\Delta U(r)&=0\,,\quad r\in \Omega_1\cup\Omega_2\,,
\label{eq:Lap1}\\
\lim_{\Omega_1\ni r\to r^\circ} U(r)&=\lim_{\Omega_2\ni r\to r^\circ} U(r)\,,
\quad r^\circ\in\Gamma\,,
\label{eq:Lap2}\\
\lim_{\Omega_1\ni r\to r^\circ}\varepsilon\nu^\circ\cdot\nabla U(r)&=
\lim_{\Omega_2\ni r\to r^\circ}        \nu^\circ\cdot\nabla U(r)\,,
\quad r^\circ\in\Gamma\,,
\label{eq:Lap3}\\
\lim_{|r|\to\infty}U(r)&=0\,,\quad r\in \Omega_1\,,
\label{eq:Lap4}
\end{align}
and they are important in the analysis of cross sections of objects
$\Omega_2$ that are much smaller than the wavelength $2\pi/k_1$.

It follows from~(\ref{eq:Lap1})--(\ref{eq:Lap4}) that the eigenfields
have zero electric energy when $\Im{\rm m}\{\varepsilon\}=0$, that is,
\begin{equation}
\int_{\Omega_1}\lvert\nabla(U(r))\rvert^2\,{\rm d}S+
\frac{1}{\varepsilon}
\int_{\Omega_2}\lvert\nabla(U(r))\rvert^2\,{\rm d}S=0\,,
\label{eq:zeroE}
\end{equation}
where ${\rm d}S$ is an element of area.

An alternative and, perhaps, more common analysis uses the scalar
electric potential $V(r)$, related to ${\bf\myvec E}(r)$
of~(\ref{eq:E}) via ${\bf\myvec E}(r)=-\nabla V(r)$. The eigenproblem
for $V(r)$ is the same as that for $U(r)$ but with $\varepsilon$
replaced by $1/\varepsilon$ in the quasi-static boundary
condition~(\ref{eq:Lap3}) and in~(\ref{eq:zeroE}). Since $U(r)$ and
$V(r)$ give rise to the same electric field we have
\begin{equation}
\nabla V(r)=
\left\{
\begin{array}{ll}
A\nabla U(r)\times\hat{\myvec z}\,,
&r\in\Omega_1\,,\\
A\varepsilon^{-1}\nabla U(r)\times\hat{\myvec z}\,,
&r\in\Omega_2\,,
\end{array}
\right.
\end{equation}
where $A$ is a normalization constant. The quasi-static potentials
$U(r)$ and $V(r)$ are still complex representations of physical fields
via~(\ref{eq:H}) and $V(r,t)=\Re{\rm e}\left\{V(r)e^{-{\rm
      i}t}\right\}$.

\subsection{Eigenfields of the semi-infinite wedge}

Magnetic and electric eigenfields can, in the limit $k_1\to 0$, only
occur when $\varepsilon$ is real and negative. When $\Gamma$ is smooth
there is a discrete set of $\varepsilon$ that admits eigenfields. When
$\Gamma$ is the open boundary of a semi-infinite wedge of opening
angle $\theta$ there exist eigenfields for all $\varepsilon$ that
satisfy~(\ref{eq:z1cont}). These wedge eigenfields
satisfy~(\ref{eq:Lap1})--(\ref{eq:Lap3}) and~(\ref{eq:zeroE}), but
not~(\ref{eq:Lap4}), and are often used in the static analysis of
singular fields in domains containing finite non-smooth objects.
See~\cite{BonnZhan17} for a rigorous justification of this practice.

The wedge eigenfields describe the singular fields that can arise at
corners of a finite object $\Omega_2$. They shed light on how incident
waves can couple very strongly to surface plasmon waves and affect
scattering and absorption cross sections of objects for all
wavenumbers $k_1$. Let $\phi$ be the azimuth angle and let $\Gamma$ be
given by $\phi=\pm\theta/2$. Then the wedge eigenfields can be found
via separation of variables and have the form
\begin{equation}
U(r)=|r|^{\pm{\rm i}\zeta}\Theta(\phi)\,,
\label{eq:Uwedge}
\end{equation}
where $\zeta$ is a real and positive parameter that solves a
transcendental equation~\cite[Section~3]{Annsop16}. The function
$\Theta(\phi)$ is odd or even depending on whether $\varepsilon<-1$ or
$\varepsilon>-1$.

The gradients $\nabla U(r)$ of the wedge eigenfields~(\ref{eq:Uwedge})
make both integrals in~(\ref{eq:zeroE}) divergent. The time-average
power loss (absorbed power) inside a disk of radius $R$, centered at
the corner vertex of the wedge, is proportional to
\begin{equation}
\lim_{\Im{\rm m}\{\varepsilon\}\to 0^+}\Im{\rm m}\{\varepsilon\}
\int_{S_{\rm w}}\lvert\nabla(U(r))\rvert^2\,{\rm d}S\,,
\label{eq:loss}
\end{equation}
where $S_{\rm w}$ is the part of the disk that overlaps the wedge. The
limit in~(\ref{eq:loss}) is non-zero and independent of $R>0$. This
means that power is absorbed by the wedge even though it becomes
lossless as $\Im{\rm m}\{\varepsilon\}\to 0^+$. A quasi-static
analysis shows that Poynting's theorem~\cite[Section~1.3.3]{Krist16}
is satisfied. The power absorption is located in a disk with an
infinitesimally small $R$ and is equal to the time-average power flow
through the boundary of a disk with arbitrarily large $R$. A detailed
analysis shows that, for the wedge eigenfields in~(\ref{eq:Uwedge}),
this power absorption depends strongly on $\varepsilon$. It goes to
zero as $\varepsilon\to -1$, but is otherwise positive.

\subsection{Coupling to surface plasmon waves}
\label{sec:coupling}

The plane wave excitation of surface plasmon waves along a boundary
$\Gamma$ with a corner can, heuristically, be explained as follows:
$U^{\rm in}(r)$ induces a magnetic field in the vicinity of the corner
that we refer to as a {\it corner field}. The corner field resembles
the wedge eigenfields~(\ref{eq:Uwedge}). According to~(\ref{eq:H}),
the time-harmonic magnetic wedge eigenfield associated
with~(\ref{eq:Uwedge}) is
\begin{equation}
H(r,t)=\Theta(\phi)\cos(\zeta\log|r|\pm t)\,.
\label{eq:magtime}
\end{equation}
This eigenfield can be viewed as a wave that travels radially inwards
($+t$) or outwards ($-t$) with a local wavelength, $2\pi|r|/\zeta$,
that increases linearly with $|r|$. The corresponding time-harmonic
corner field inherits these characteristics. At the value of $|r|$
where the local wavelength equals the wavelength of the surface
plasmon waves, $2\pi/k_{\rm sp}$ see~(\ref{eq:spl}), the corner field
couples to the surface plasmon waves on $\Gamma$. Only odd corner
fields can couple to surface plasmon waves since the latter only exist
for $\varepsilon<-1$, see Section~\ref{sec:Drude}.

\section{Cross sections}
\label{sec:cross}

With the incident plane wave $U^{\rm in}(r)$ of~(\ref{eq:Uin}), the
scattering cross section, $\sigma_{\rm sc}$, and the absorption cross
section, $\sigma_{\rm abs}$, of $\Omega_2$ are defined as time
averages of the scattered and absorbed power densities divided by the
time average of the incident power density. Let $\Gamma_{\rm C}$ be a
contour enclosing $\Omega_2$ and with outward unit normal $\nu$.
Then~\cite[Section~4.2]{Krist16}
\begin{align}
\sigma_{\rm sc}&=\Im{\rm m}\left\{\frac{1}{k_1}\int_{\Gamma_{\rm C}}
\left(\nu\cdot\nabla U^{\rm sc}(r)\right)\overline{U}^{\rm sc}(r)\,{\rm d}\ell
\right\},
\label{eq:sca}\\
\sigma_{\rm abs}&=-\Im{\rm m}\left\{\frac{1}{k_1}\int_{\Gamma_{\rm C}}
\left(\nu\cdot\nabla U(r)\right)\overline{U}(r)\,{\rm d}\ell
\right\}.
\label{eq:abs}
\end{align}
When the object has corners and the real part of $\varepsilon$ is
negative, the absorption cross section can be positive, even in the
limit of the imaginary part of $\varepsilon$ going to zero. The
absorption can be explained by the quasi-static analysis in
Section~\ref{sec:sing} and is verified numerically in
Section~\ref{sec:absorp}.

The total cross section in the direction $d$ of $U^{\rm in}(r)$ is
\begin{equation}
\sigma_{\rm tot}=\sigma_{\rm sc}+\sigma_{\rm abs}\,.
\label{eq:stot1}
\end{equation}
The optical theorem,
\cite[Section~10.11]{Jackson1999},~\cite[Section~4.4]{Krist16}, gives
the alternative expression for the total cross section
\begin{equation}
\sigma_{\rm tot}=-\lim_{|r|\to\infty}\Im{\rm m}\left\{
\frac{4}{k_1}U^{\rm sc}(|r|d)\sqrt{\frac{\pi k_1|r|}{2}}
e^{-{\rm i}(k_1|r|-\pi/4)}
\right\}\,.
\label{eq:stot2}
\end{equation}

\section{Numerical examples}
\label{sec:numex}

In a series of progressively more challenging problems we now put the
integral equations of Section~\ref{sec:inteq} and the discretization
techniques of Section~\ref{sec:disc} to the test. Only a few of our
problems have (semi-)analytic solutions. When assessing the accuracy
of computed quantities we therefore often adopt a procedure where to
each numerical solution we also compute an overresolved reference
solution, using roughly 50\% more points in the discretization of the
integral equations. The absolute difference between these two
solutions is denoted the {\it estimated absolute error}.

Our codes are implemented in {\sc Matlab}, release 2016b, and executed
on a workstation equipped with an Intel Core i7-3930K CPU. The
implementations are standard, rely on built-in functions, and include
a few {\tt parfor}-loops (which execute in parallel).

\subsection{Numerical tests of integral operators}

The operators $S_k$, $K_k$, $K^{\rm A}_k$ and $T_k$, $r\in\Gamma$,
have been implemented on the ``star'' boundary~\cite{Hao14,HelsHols15}
parameterized as
\begin{equation}
r(s)=\frac{9}{20}\left(1+\frac{20}{81}\sin(5s)\right)(\cos(s),\sin(s))\,,
\quad -\pi\le s\le\pi\,.
\label{eq:star}
\end{equation}
Product integration weights for kernels with logarithmic singularities
are computed using analytical methods and
recursion~\cite[Appendix~A]{HelsHols15}, while local
regularization~\cite[Section~2.2]{Hels09} is used for hypersingular
kernels.

The compositions of operators $K_kK_k-S_kT_k$ and $K^{\rm A}_kK^{\rm
  A}_k-T_kS_k$ act as the identity operator on simple smooth layer
densities, compare~(\ref{eq:cald1}) and~(\ref{eq:cald2}). For example,
with $k=3.8+1.3{\rm i}$, $f(r(s))=\cos(3s)+{\rm i}\sin(7s)$, and 384
discretization points on $\Gamma$ of~(\ref{eq:star}), the relation
$(K_kK_k-S_kT_k)f(r)=f(r)$ holds with a relative accuracy of $4\cdot
10^{-15}$ in $L^2$-norm. With 1152 discretization points, the relation
$(K^{\rm A}_kK^{\rm A}_k-T_kS_k)f(r)=f(r)$ holds with a relative
accuracy of $4\cdot 10^{-14}$.

\subsection{Eigenwavenumbers for the unit circle}

We first choose $\varepsilon=2.25$ so that $\arg(k_1)=\arg(k_2)$, let
$\Gamma$ be the unit circle, and look for true and false
eigenwavenumbers $k_1$ with $\Re{\rm e}\{k_1\}>0$ using 352
discretization points on $\Gamma$. We investigate the KM0 system,
which is~(\ref{eq:KM1sys}) with $c=1$, and the KM2
system~(\ref{eq:KM2sys}). See Section~\ref{sec:trufaleig} for the
definition of true and false eigenwavenumbers.

\begin{table}
\caption{Estimates of true and false eigenwavenumbers of the 
KM0 and KM2 systems for the unit circle with $\varepsilon=2.25$.}
\vspace{0.2cm}
\centering
\begin{tabular}{|c|c|c|}
\hline
 $k_1$ & multiplicity & nature  \\
\hline
 $2.380109395443269-0.303953834460040{\rm i}$ & simple & false \\
 $3.041565475205771-1.041465761622153{\rm i}$ & double & true  \\
 $3.815540575399378-0.309076450175921{\rm i}$ & double & false \\
 $4.892032383544720-0.631231166352111{\rm i}$ & double & true  \\
\hline
\end{tabular}
\label{tab:k1ex}
\end{table}

\begin{figure}
\centering 
  \includegraphics[height=55mm]{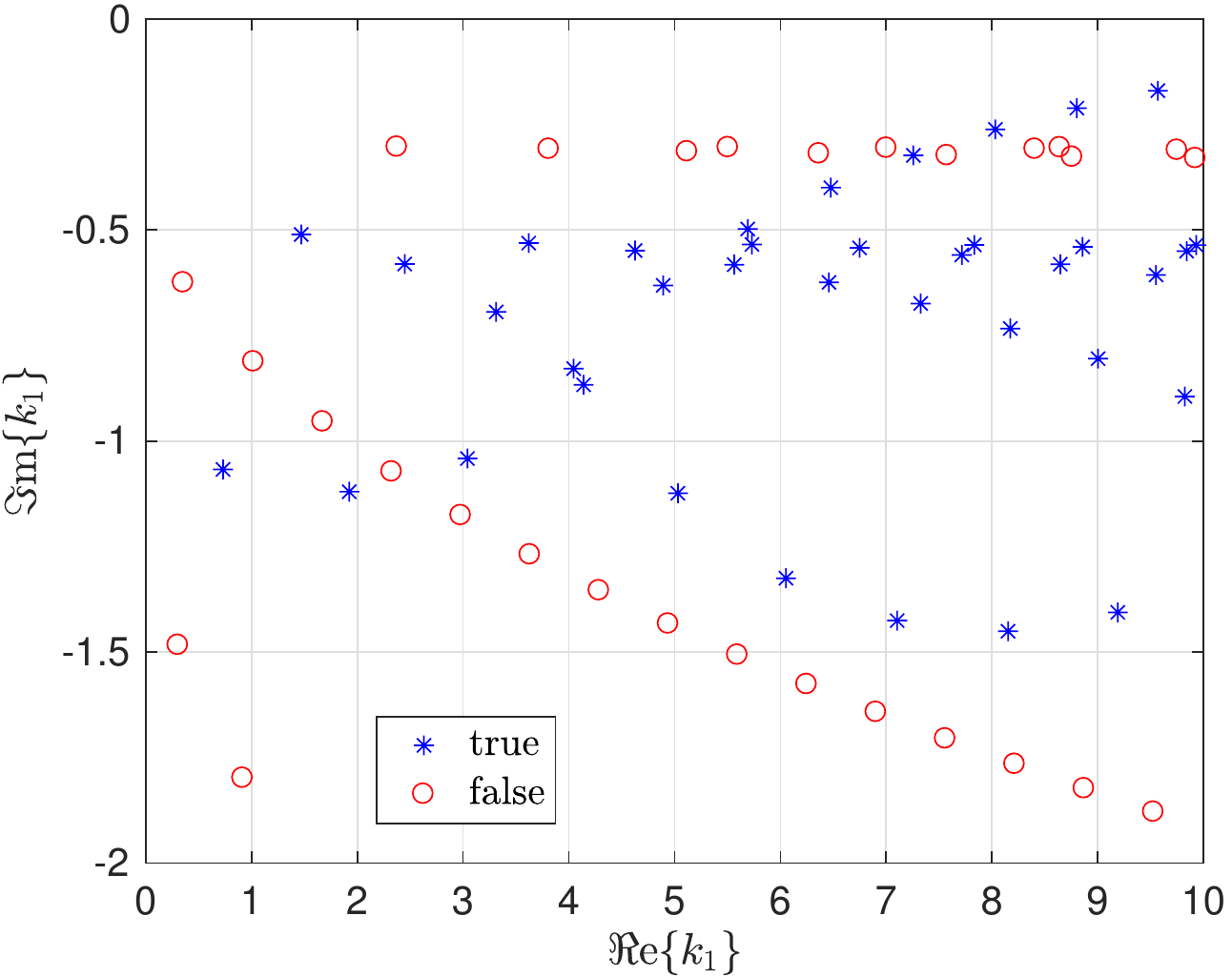}
\caption{\sf True and false eigenwavenumbers $k_1$ of the 
  KM0 and KM2 systems for the unit circle with $\varepsilon=2.25$.}
\label{fig:mreig}
\end{figure}

As it turns out in our numerical experiments, the eigenwavenumbers of
the KM0 system and those of the KM2 system are the same. A few
examples are listed in Table~\ref{tab:k1ex}. The true eigenwavenumbers
are confirmed to a relative precision of $4\cdot 10^{-16}$ by
comparison with semi-analytic results, computed as solutions to
transcendental equations derived in analogy with their
three-dimensional counterparts in~\cite[Section~3]{KresRoac78}. We
believe that this precision is indicative of the precision in all
computed eigenwavenumbers in this and the following sections.
Figure~\ref{fig:mreig} illustrates all eigenwavenumbers found with
$0\le\Re{\rm e}\{k_1\}\le 10$ and $\Im{\rm m}\{k_1\}\ge -2$. The
eigenwavenumbers are found using a combination of brute-force random
search and Broyden's method, see~\cite[Section~VI.B]{HelsKarl17} for a
few more details. No eigenwavenumber has $\Im{\rm m}\{k_1\}\ge0$, in
agreement with the theory of Section~\ref{sec:KM1unique}.

\begin{figure}[t]
\centering 
  \includegraphics[height=47mm]{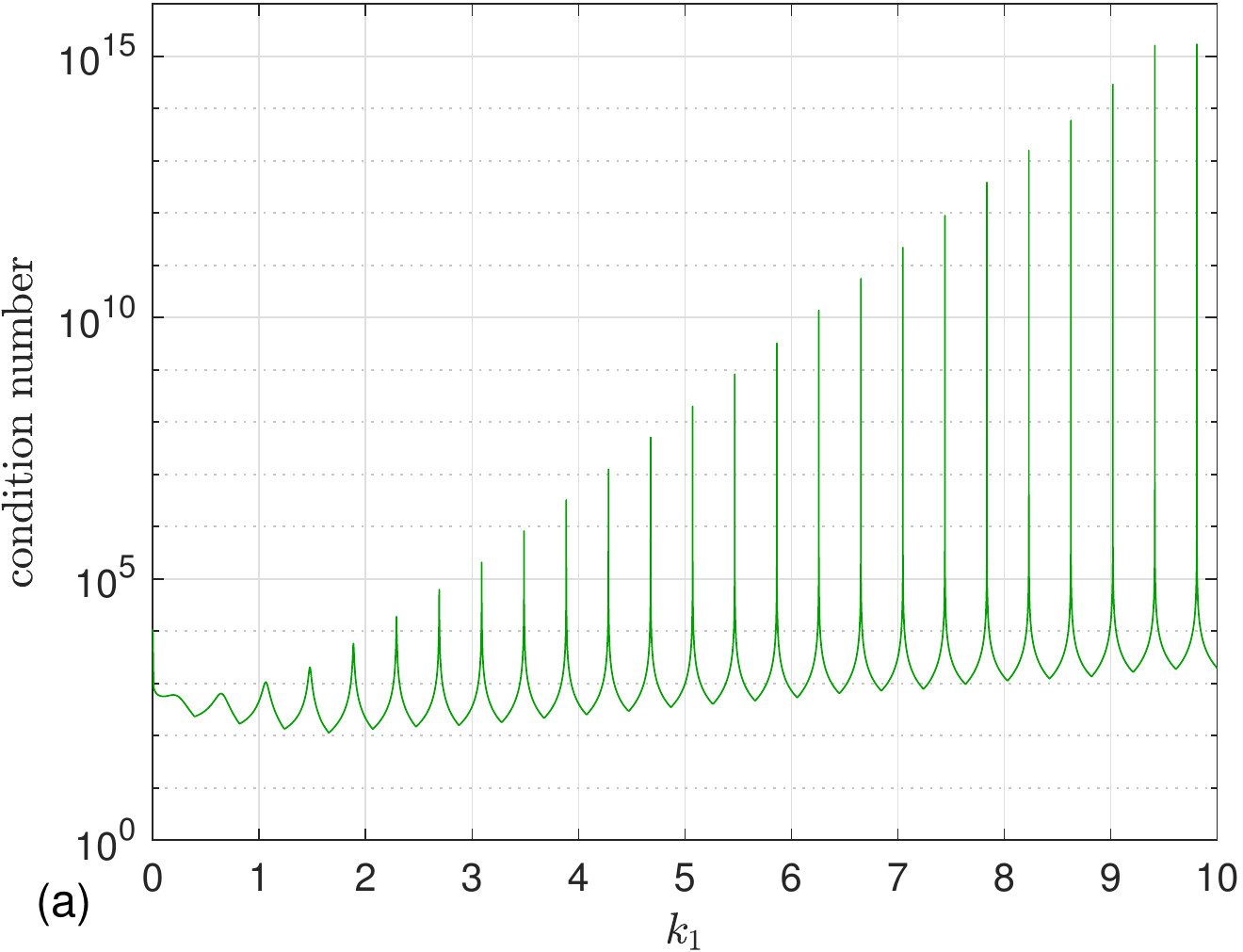}
  \includegraphics[height=47mm]{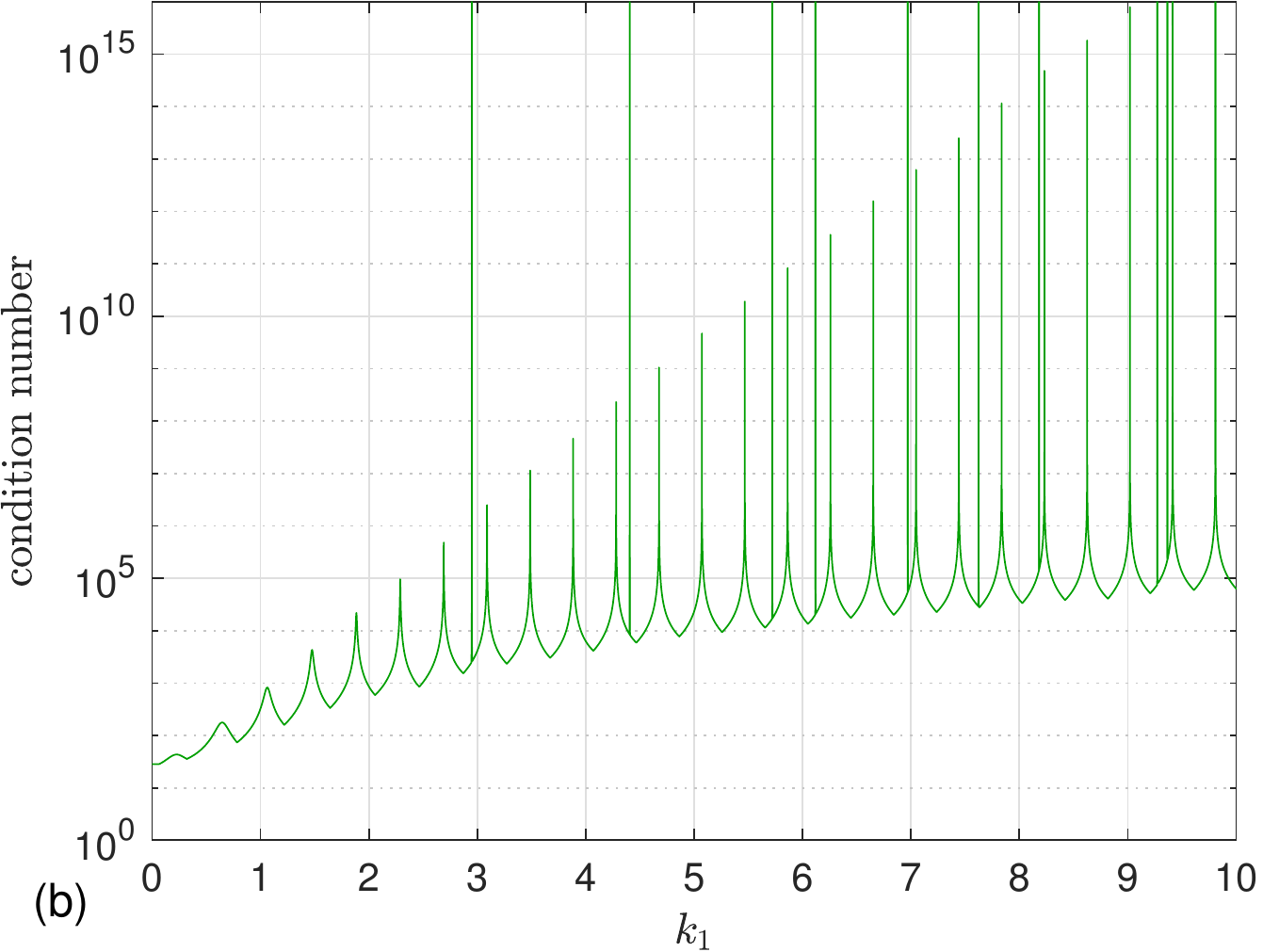}
\caption{\sf Condition numbers of system matrices from the KM1 and
  the KM0 systems for the unit circle, $\varepsilon=-1.1838$, and
  $k_1\in[0,10]$: (a) the KM1 system is free of false
  eigenwavenumbers; (b) the KM0 system exhibits nine false
  eigenwavenumbers.}
\label{fig:cond}
\end{figure}

We then choose $\varepsilon=-1.1838$, which is used
in~\cite{Annsop16}. The condition numbers of the matrices resulting
from discretization of the KM1 system and the KM0 system are studied
as a function of $k_1\in[0,10]$ using 384 discretization points on
$\Gamma$. Note that KM1, according to~(\ref{eq:cchoice}), now has
$c=-{\rm i}$ while KM0 always corresponds to $c=1$, so the two systems
are not the same. Figure~\ref{fig:cond}(a) shows that the KM1 system
does not exhibit any false eigenwavenumber, in agreement with the
theory of Section~\ref{sec:KM1unique}. The KM0 system exhibits nine
false eigenwavenumbers, see Figure~\ref{fig:cond}(b). Furthermore, the
KM1 system leads to generally better conditioned matrices. The sharp
peaks that are common to Figures~\ref{fig:cond}(a)
and~\ref{fig:cond}(b) are caused by eigenwavenumbers close to, but
below, the real $k_1$-axis. Results for the KM2 system (not shown) are
very similar to those of the KM0 system: the false eigenwavenumbers
are the same, but the condition numbers are generally slightly
smaller.

\begin{figure}
\centering 
  \includegraphics[height=51mm]{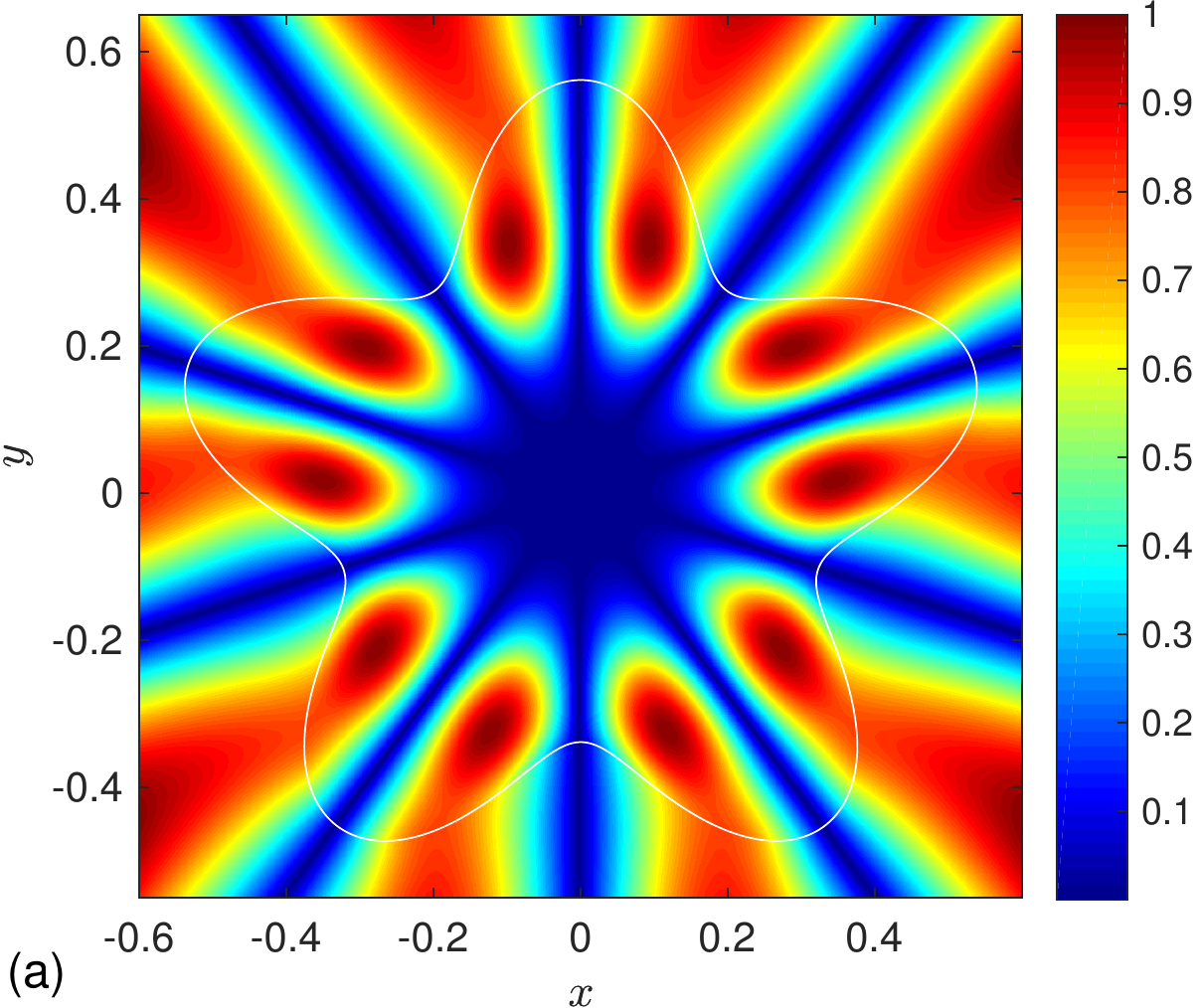}
  \includegraphics[height=51mm]{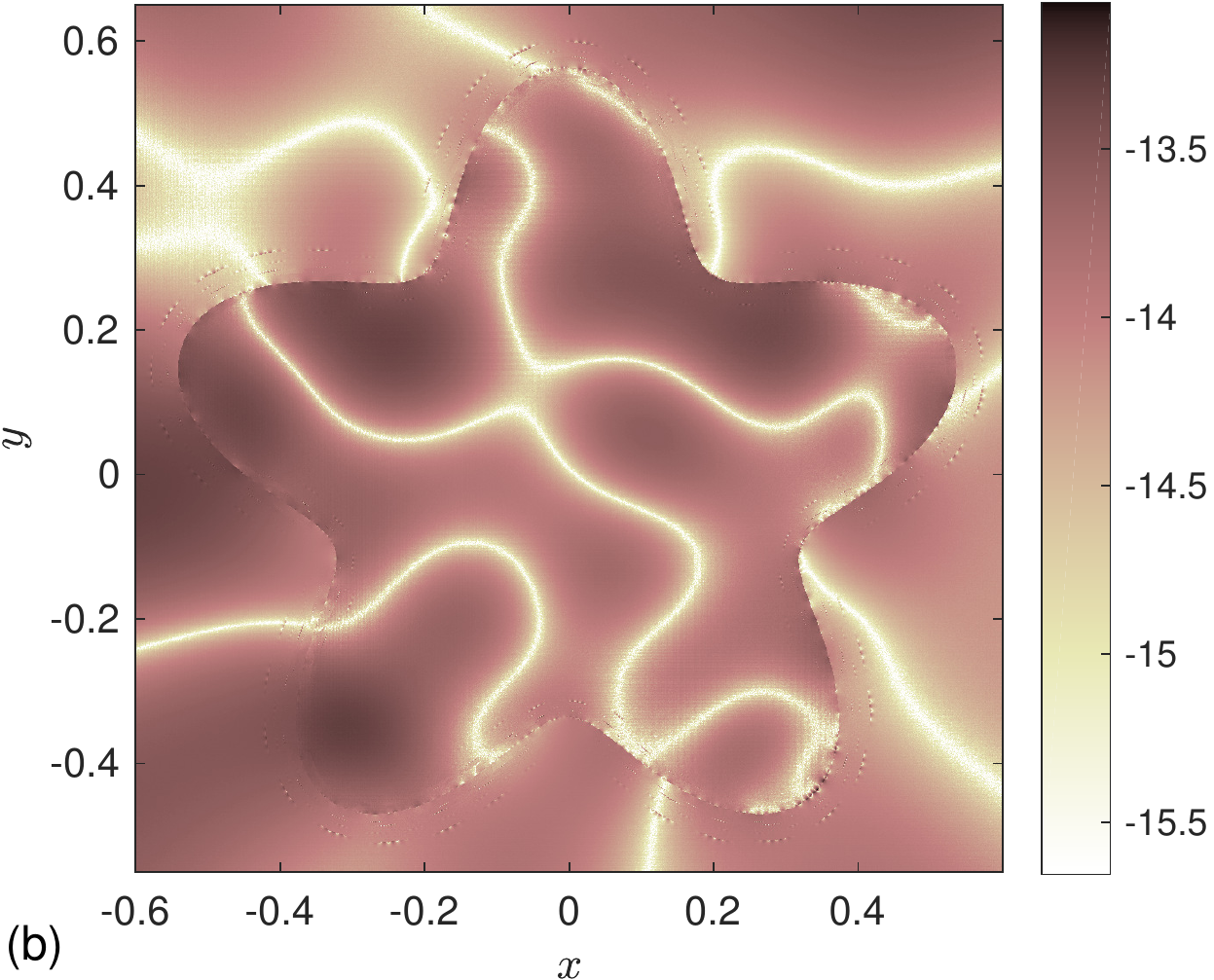}
\caption{\sf Normalized eigenfield $U(r)$ for $\Gamma$
  of~(\ref{eq:star}) and with $\varepsilon=2.25$ and estimated
  eigenwavenumber $k_1=13.21401616284493-1.636497767435982{\rm i}$:
  (a) absolute field value $|U(r)|$; (b) $\log_{10}$ of estimated
  absolute field error in $|U(r)|$.}
\label{fig:field1}
\end{figure}

\subsection{Eigenfield for a ``star''}
\label{sec:starfield}

We choose $\varepsilon=2.25$ for the ``star'' of~(\ref{eq:star}),
again look for non-trivial solutions to the homogeneous KM0 system,
and then compute eigenfields $U(r)$ via~(\ref{eq:U1KM1})
and~(\ref{eq:U2KM1}) at $10^6$ field points placed on a Cartesian grid
in the box ${\cal B}=\left\{-0.6\le x\le 0.6, -0.55\le y\le
  0.65\right\}$. The eigenfields are normalized with their largest
value in ${\cal B}$.

Figure~\ref{fig:field1} shows the eigenfield for the simple estimated
eigenwavenumber $k_1=13.21401616284493-1.636497767435982{\rm i}$ along
with the estimated absolute field-error computed with 976
discretization points on $\Gamma$. The accuracy is very high, also
close to $\Gamma$, which demonstrates the power of the near-boundary
evaluation scheme of Section~\ref{sec:disc}.

\begin{figure}
\centering 
  \includegraphics[height=47mm]{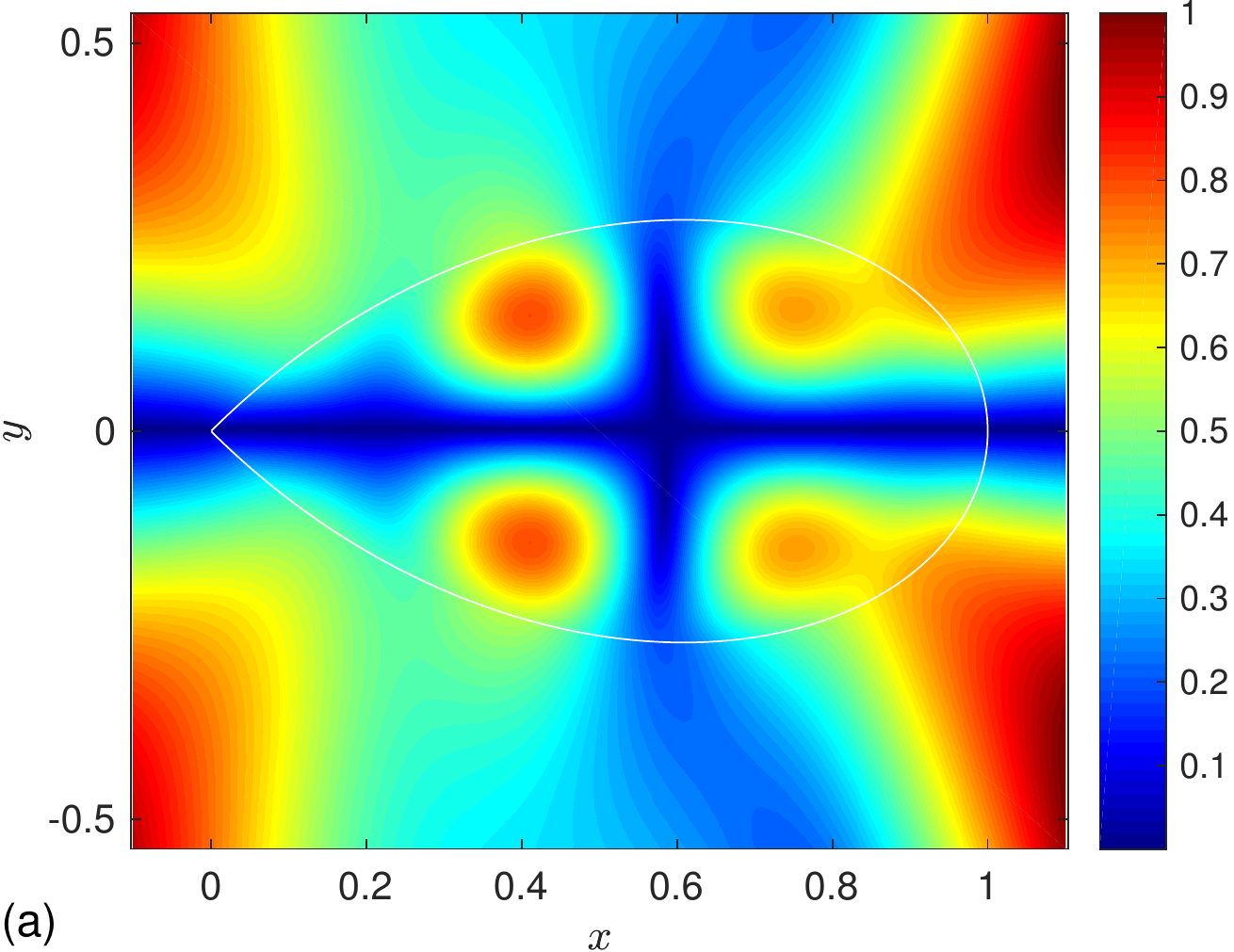}
  \includegraphics[height=47mm]{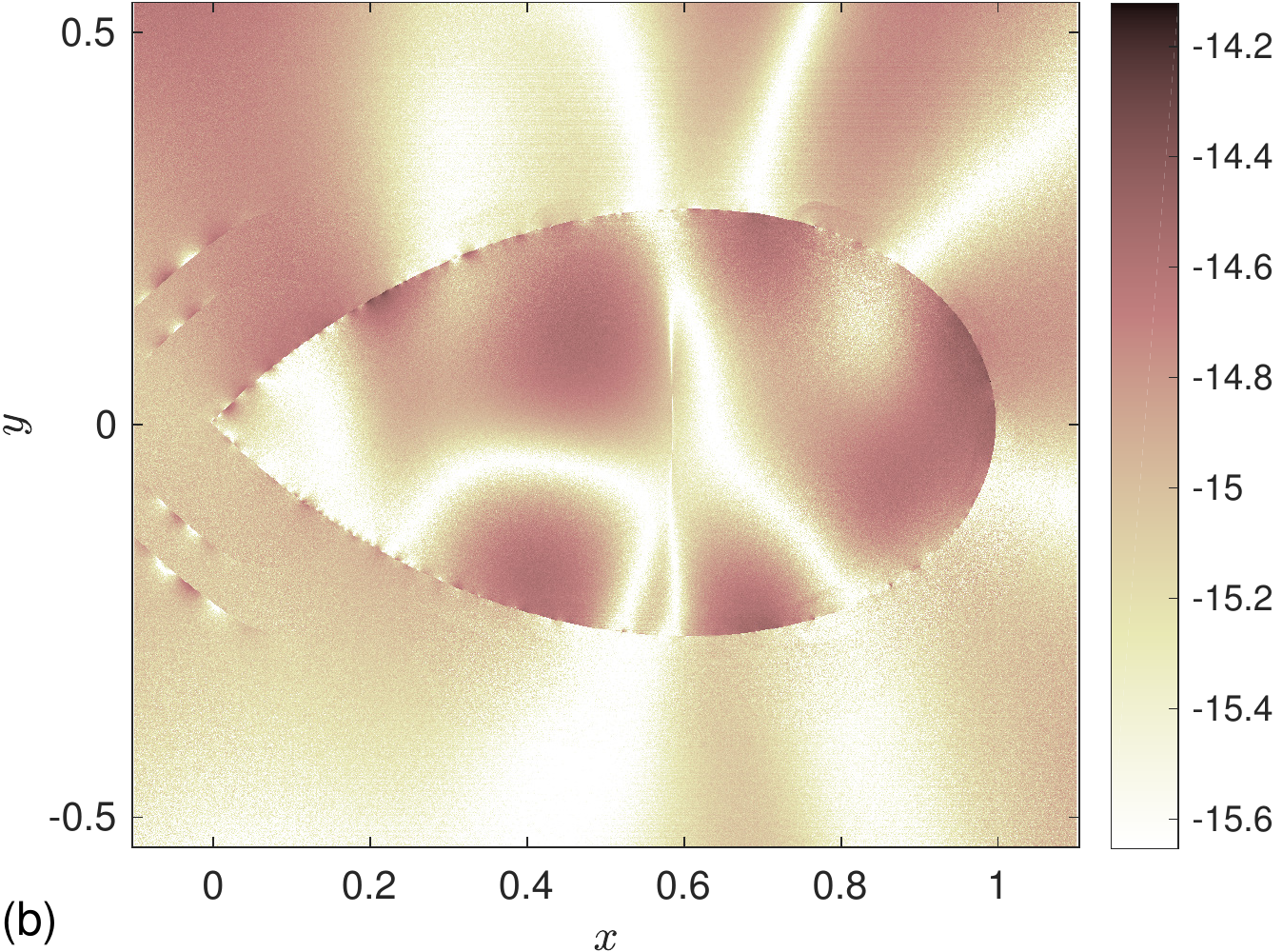}
\caption{\sf Normalized eigenfield $U(r)$ for $\Gamma$
  of~(\ref{eq:Gamma}) and with $\varepsilon=2.25$ and estimated
  eigenwavenumber $k_1=9.701129417644246-2.000374579086419{\rm i}$:
  (a) absolute field value $|U(r)|$; (b) $\log_{10}$ of estimated
  absolute field error in $|U(r)|$.}
\label{fig:field2}
\end{figure}

\subsection{Eigenfield for a one-corner object}

We now let $\Gamma$ be a closed contour with one corner, parameterized
as
\begin{equation}
r(s)=\sin(\pi s)\left(\cos((s-0.5)\theta),\sin((s-0.5)\theta)\right)\,,
\quad 0\le s\le 1\,.
\label{eq:Gamma}
\end{equation}
We choose $\varepsilon=2.25$ and $\theta=\pi/2$ and repeat the
experiment of Section~\ref{sec:starfield} with the KM0 system. RCIP
acceleration is activated due to the presence of the corner, see
Section~\ref{sec:disc}, and 320 discretization points are placed on
the coarse mesh on $\Gamma$. Figure~\ref{fig:field2} shows the
eigenfield for the simple estimated eigenwavenumber
$k_1=9.701129417644246-2.000374579086419{\rm i}$ along with the
estimated absolute error at $10^6$ field points. The accuracy is even
higher than in the example of Section~\ref{sec:starfield}, which
demonstrates the power of RCIP and that boundary value problems on
domains with corners are not necessarily more difficult to solve than
boundary value problems on smooth domains.

\subsection{Surface plasmon wave for a one-corner object}
\label{sec:spw}

We solve the KM1 equations with $\varepsilon=-1.1838$, $k_1=18$,
$\Gamma$ as in~(\ref{eq:Gamma}), $\theta=\pi/6$, and $U^{\rm in}(r)$
as in~(\ref{eq:Uin}) with $d=(\cos(5\pi/12),\sin(5\pi/12))$. This
setup is chosen as to create a surface plasmon wave and to resemble
the setup of~\cite[Section~4.4.1]{Annsop16}.

\begin{figure}
\centering 
  \includegraphics[height=47mm]{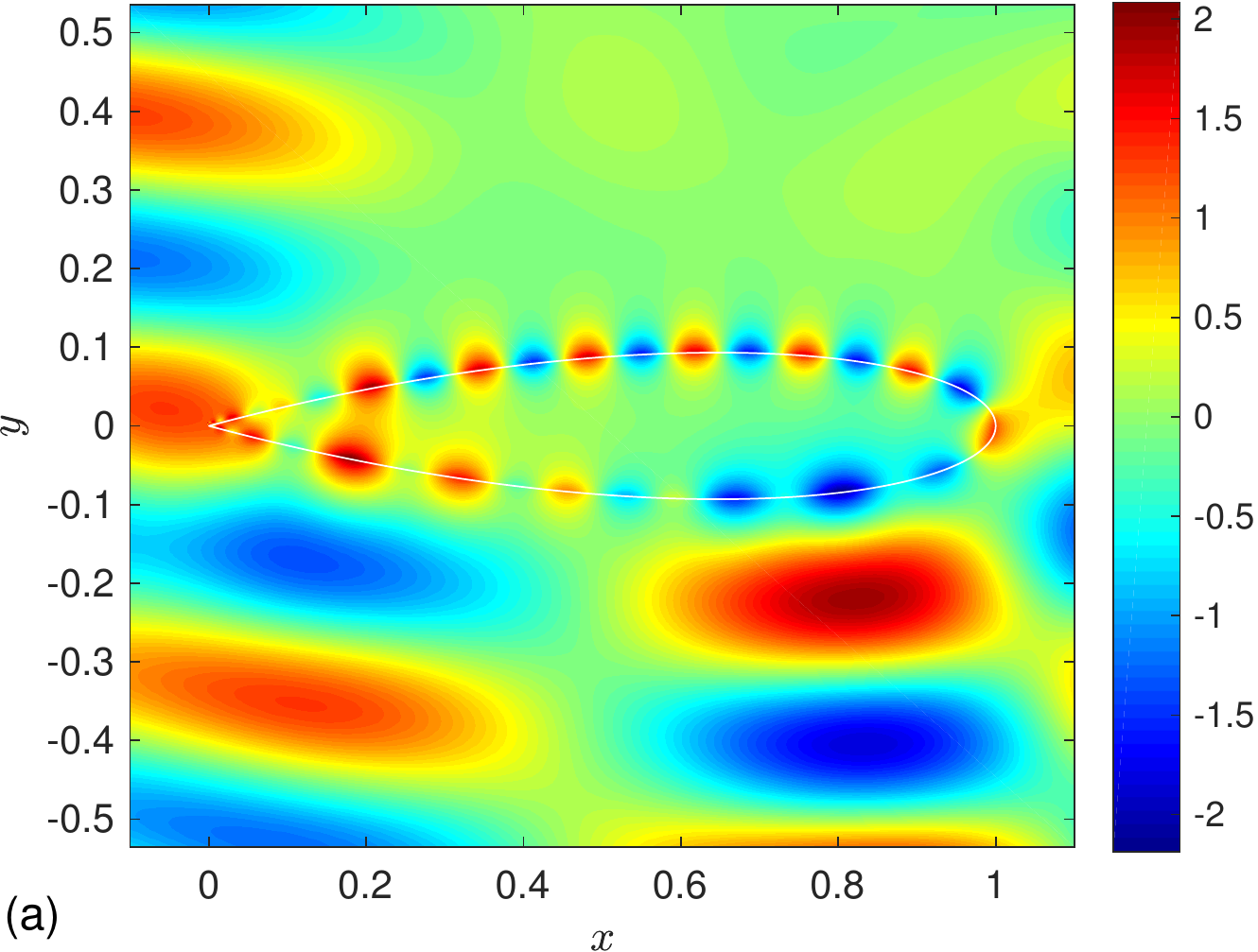}
  \includegraphics[height=47mm]{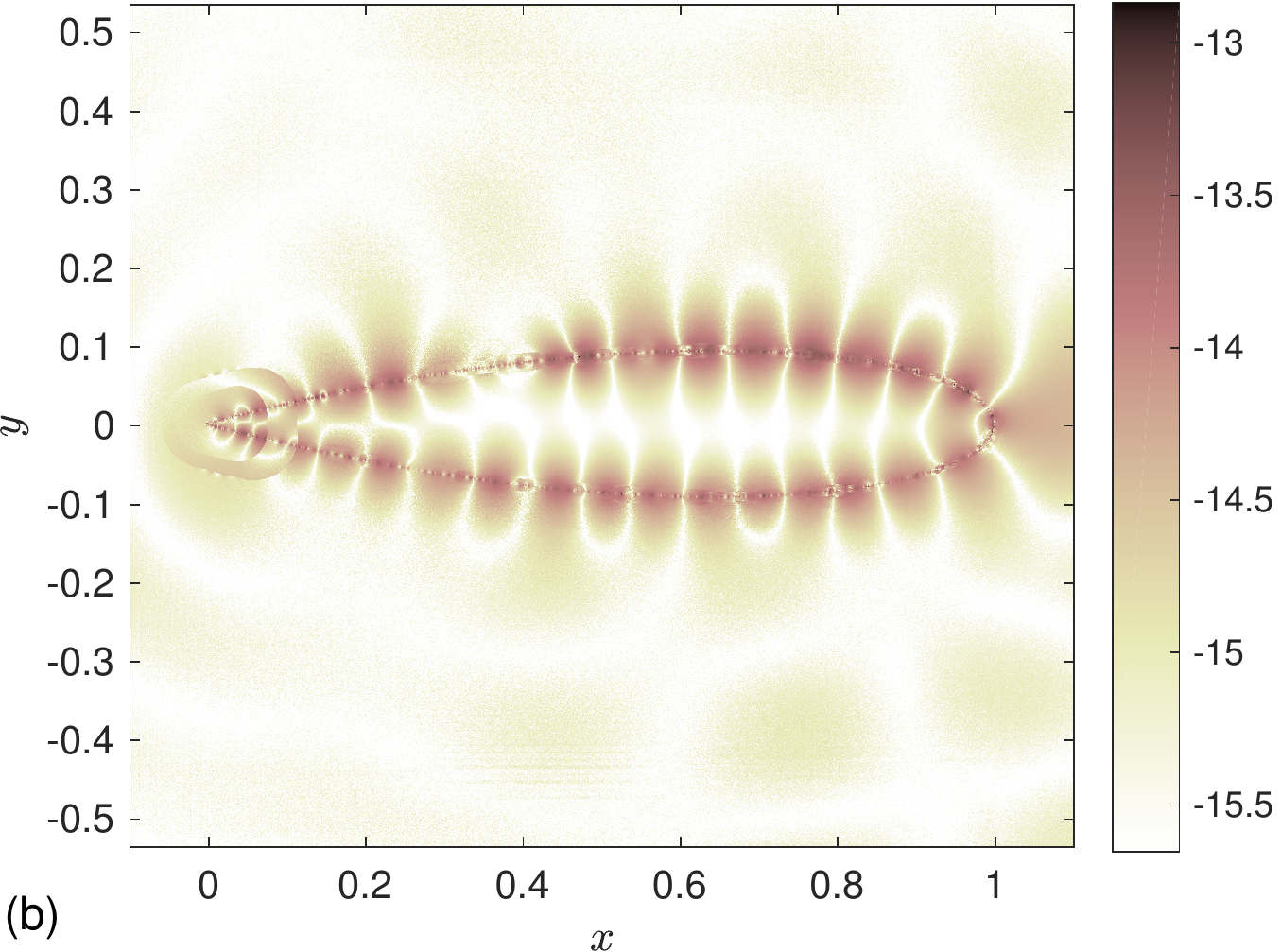}
  \includegraphics[height=46mm]{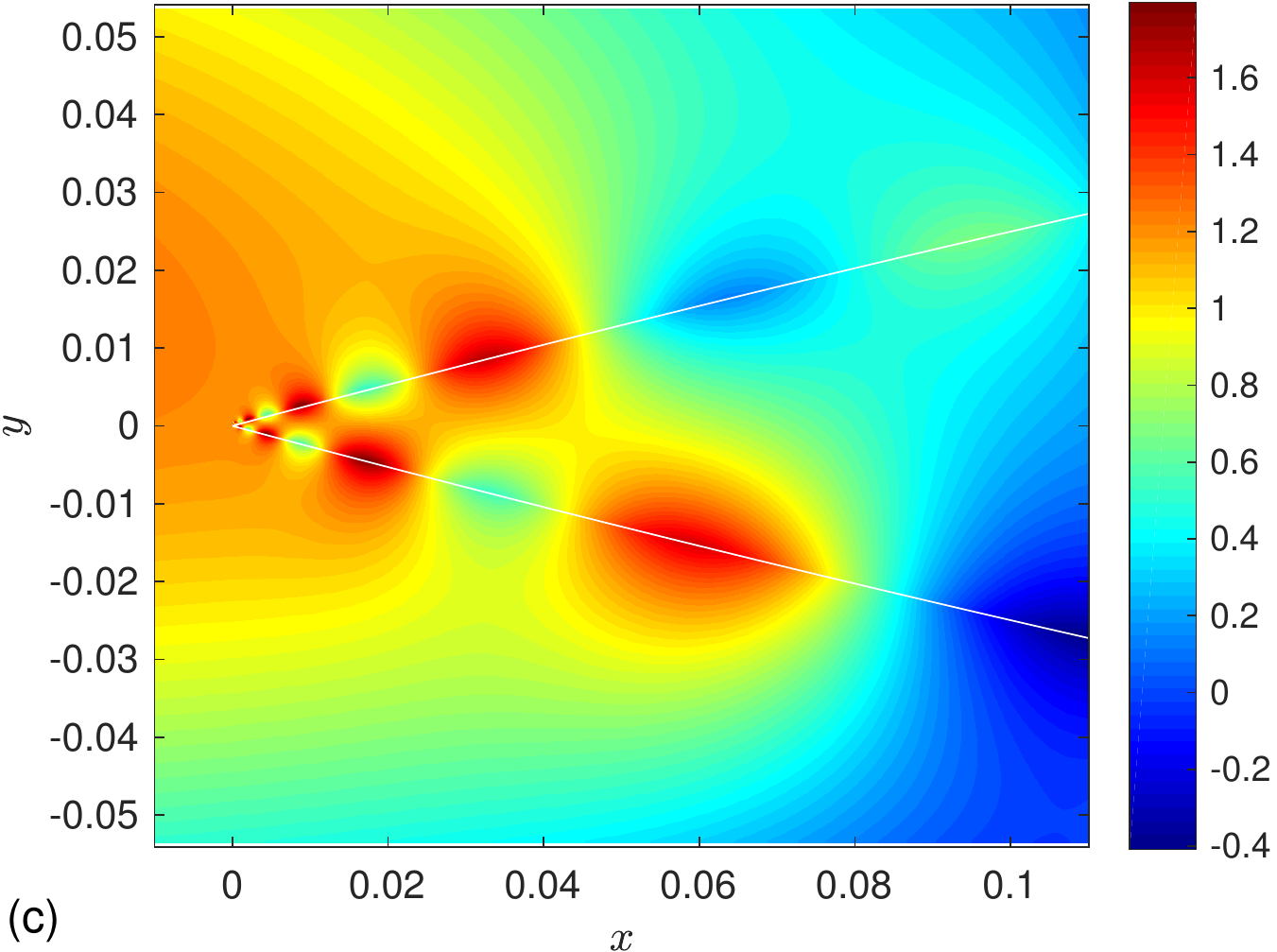}
  \includegraphics[height=46mm]{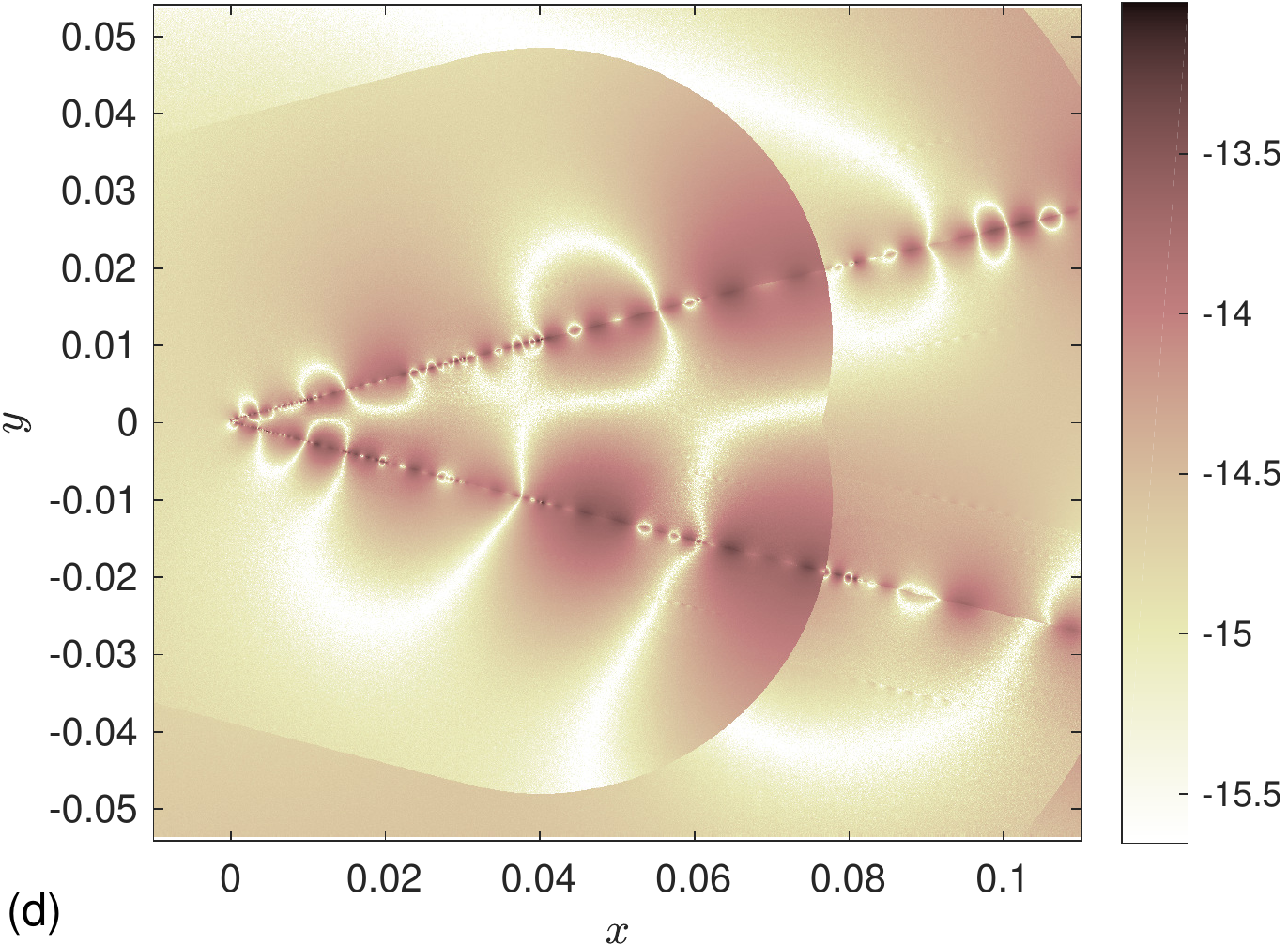}
  \includegraphics[height=49mm]{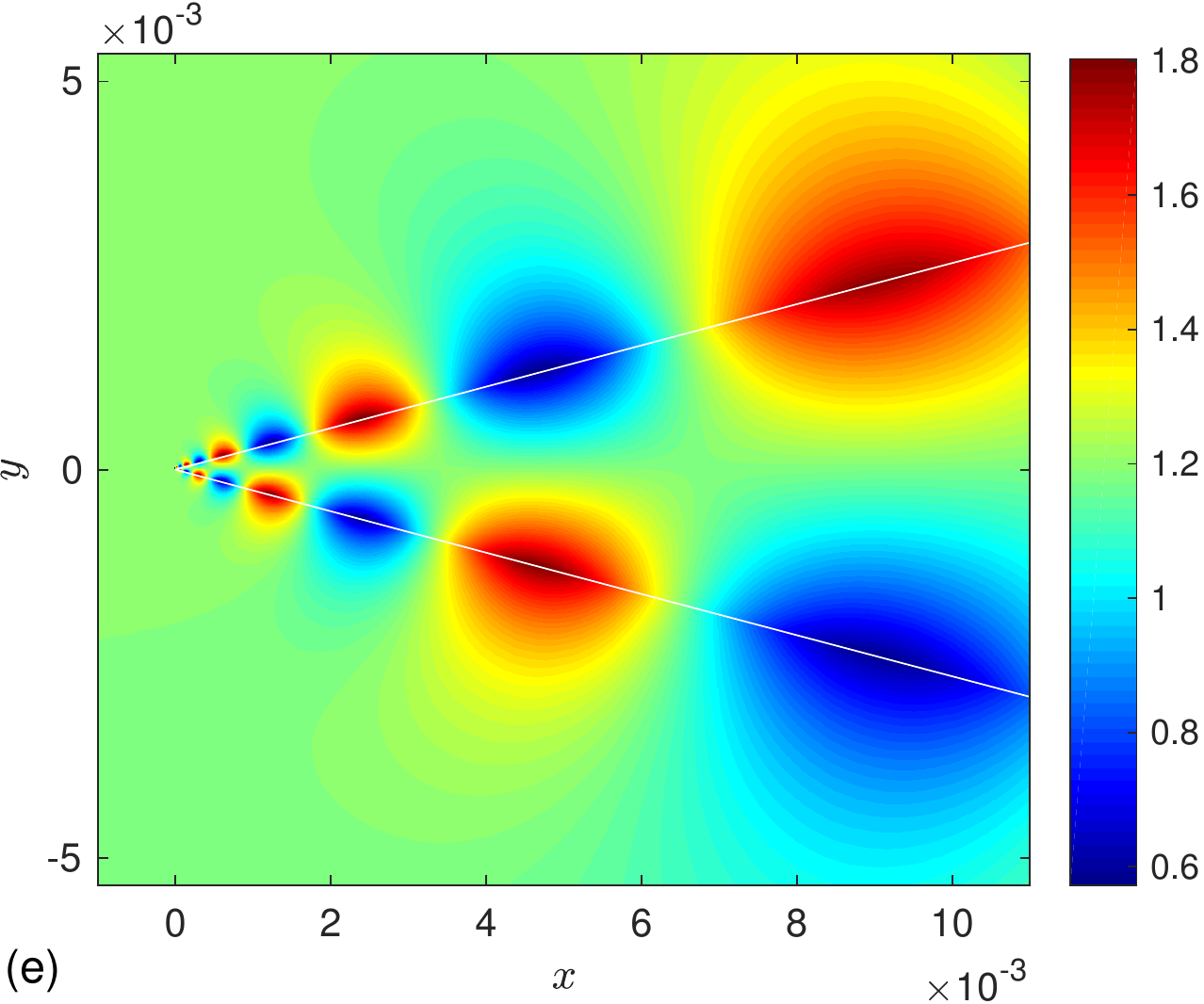}
  \hspace*{1mm}
  \includegraphics[height=49mm]{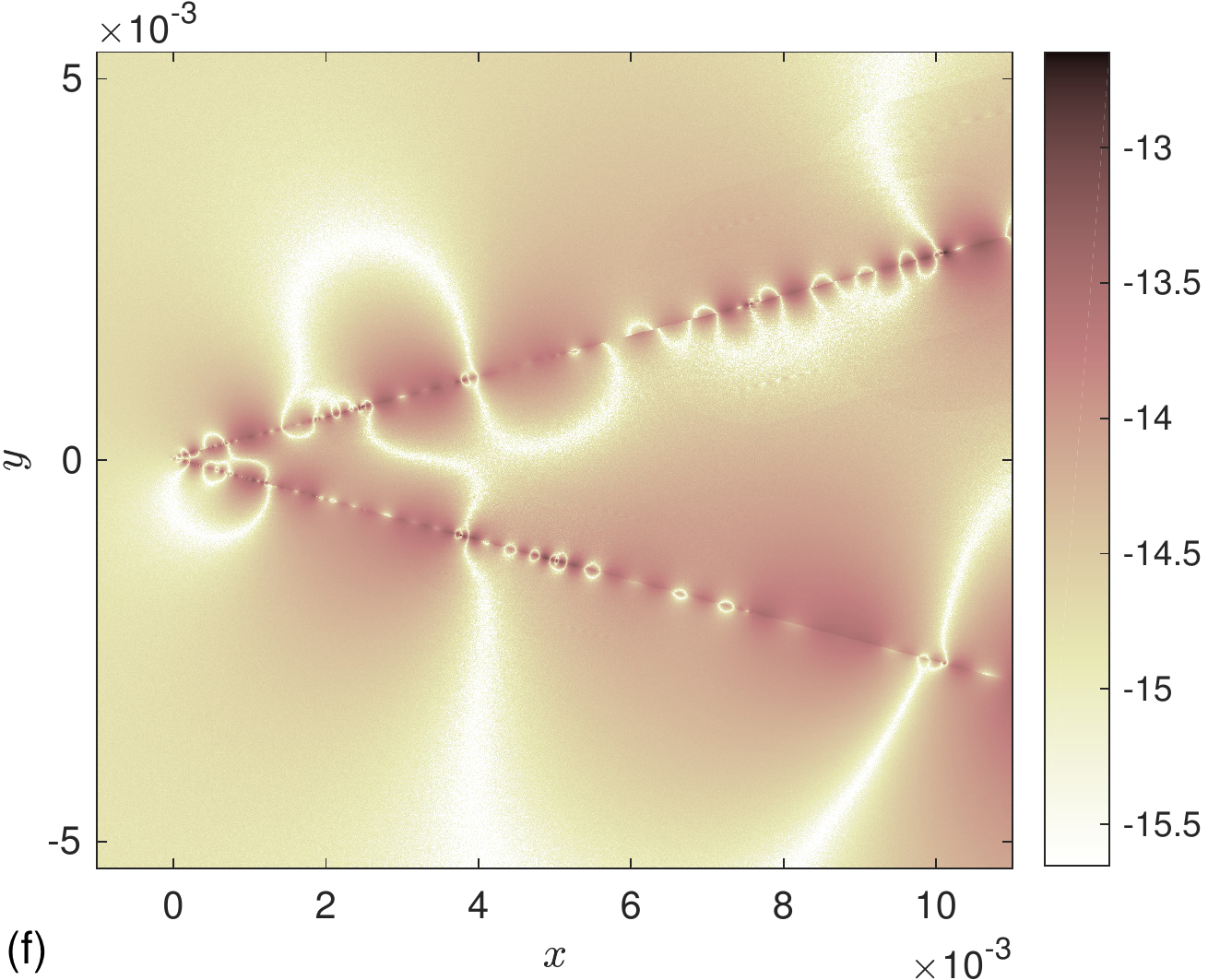}
\caption{\sf A corner zoom for $\varepsilon=-1.1838$, $\Gamma$ as 
  in~(\ref{eq:Gamma}), $\theta=\pi/6$, $k_1=18$, and
  $d=\left(\cos(5\pi/12),\sin(5\pi/12)\right)$: (a,b) $H^+(r,0)$
  of~(\ref{eq:H}) and $\log_{10}$ of estimated absolute error; (c,d)
  ten times magnification; (e,f) 100 times magnification.}
\label{fig:plzoom}
\end{figure}

The KM1 system~(\ref{eq:KM1sysc}), which now has $c=-{\rm i}$
according to~(\ref{eq:cchoice}), is solved for the limit solution
$\mu^+$, $\rho^+$. The representations~(\ref{eq:U1KM1})
and~(\ref{eq:U2KM1}) are used for $U^+(r)$ and~(\ref{eq:H}) is used
for $H^+(r,0)$. Some resolution issues, related to the relatively
small opening angle $\theta=\pi/6$ and quadrature panels on opposing
sides of the corner vertex lying close to each other, require that the
number of points used in the underlying quadrature is increased from
$n_{\rm pt}=16$ to $n_{\rm pt}=22$.
Compare~\cite[Section~21.1]{Hels17}.

Figure~\ref{fig:plzoom} shows a sequence of zooms of $H^+(r,0)$ in the
vicinity of the corner vertex, along with a plot of the estimated
absolute field error. There are 1100 discretization points on the
coarse mesh on $\Gamma$ and each computational box contains $10^6$
field points on a (rectangular) Cartesian grid. According to the
analysis of Section~\ref{sec:Drude}, a surface plasmon wave can
propagate along $\Gamma$ with a wavelength $2\pi/k_{\rm sp}\approx
0.138$. Figure~\ref{fig:plzoom}(a) shows that this indeed happens. An
animation of a surface plasmon wave $H^+(r,t)$, $t\in[0,2\pi]$, along
$\Gamma$ can be found in~\cite{anim}.

The estimated field accuracy is not affected by the proximity of a
field point to the corner vertex, as shown in
Figures~\ref{fig:plzoom}(b,d,f). At least thirteen digits can be
obtained irrespective of the level of zoom.
Figure~\ref{fig:plzoom}(a,c,e) can serve as an illustration to the
discussion in Section~\ref{sec:coupling} of how the odd magnetic
eigenfields~(\ref{eq:magtime}) couple to the surface plasmon waves.

\begin{figure}
\centering 
  \includegraphics[height=48mm]{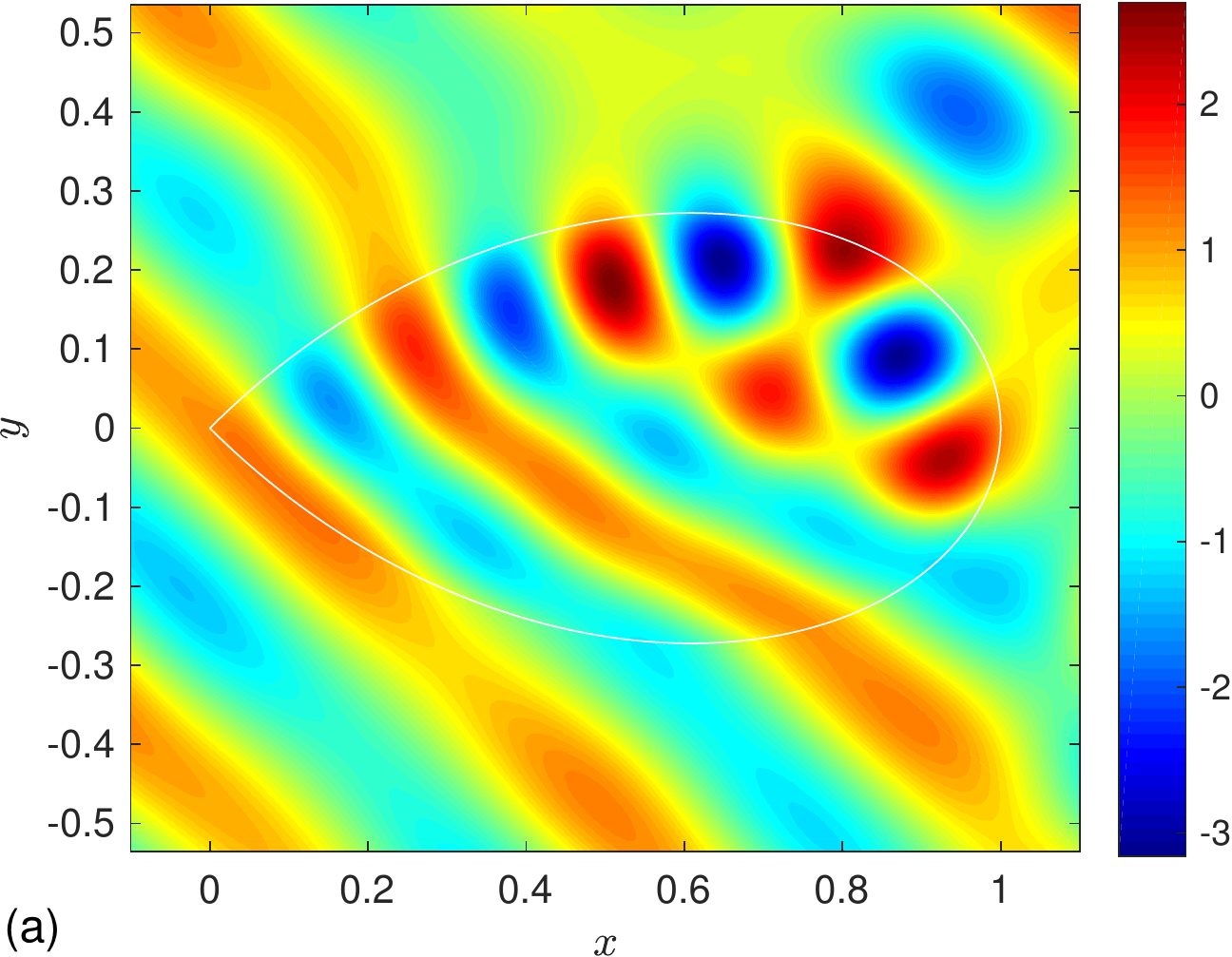}
  \includegraphics[height=48mm]{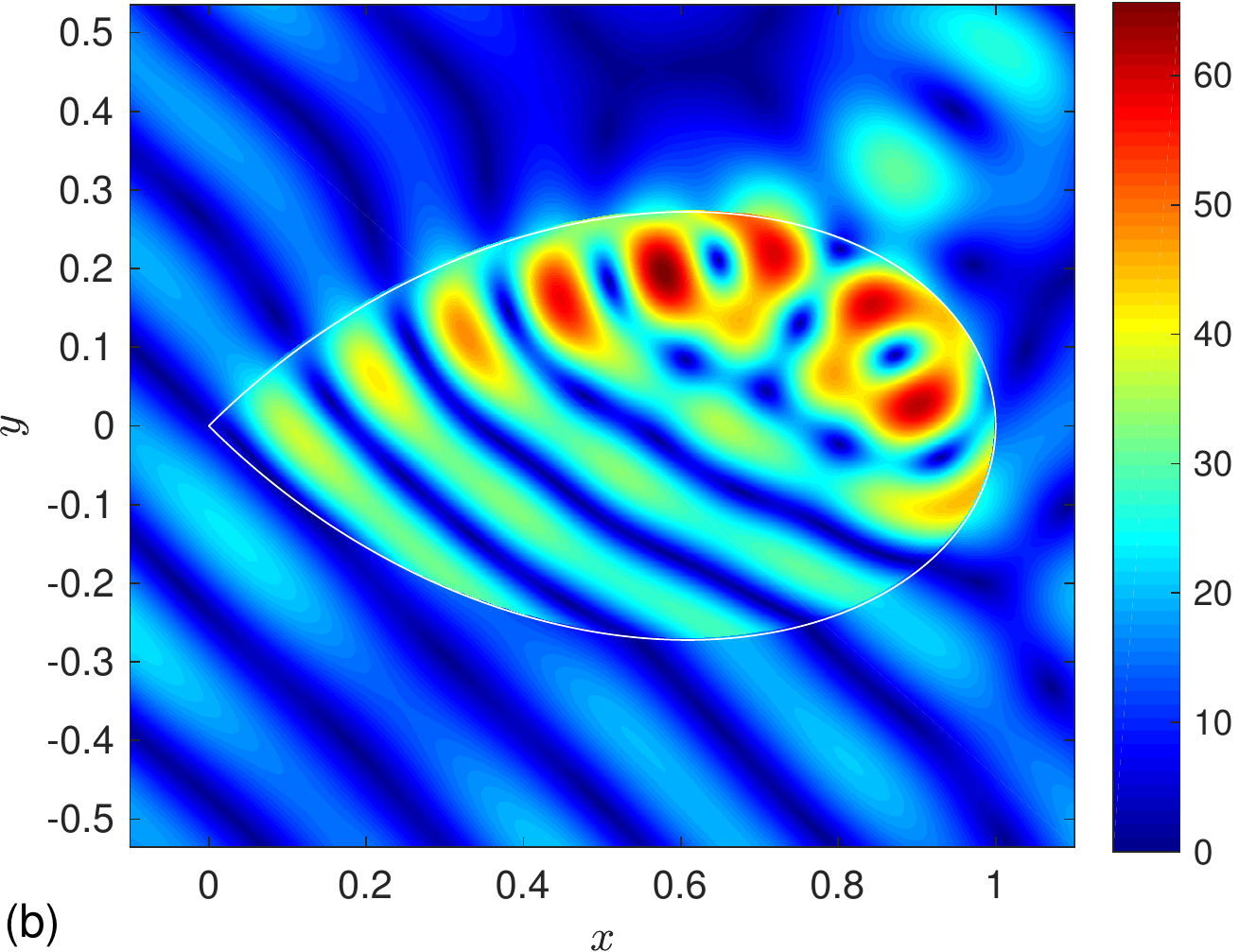}
  \includegraphics[height=47mm]{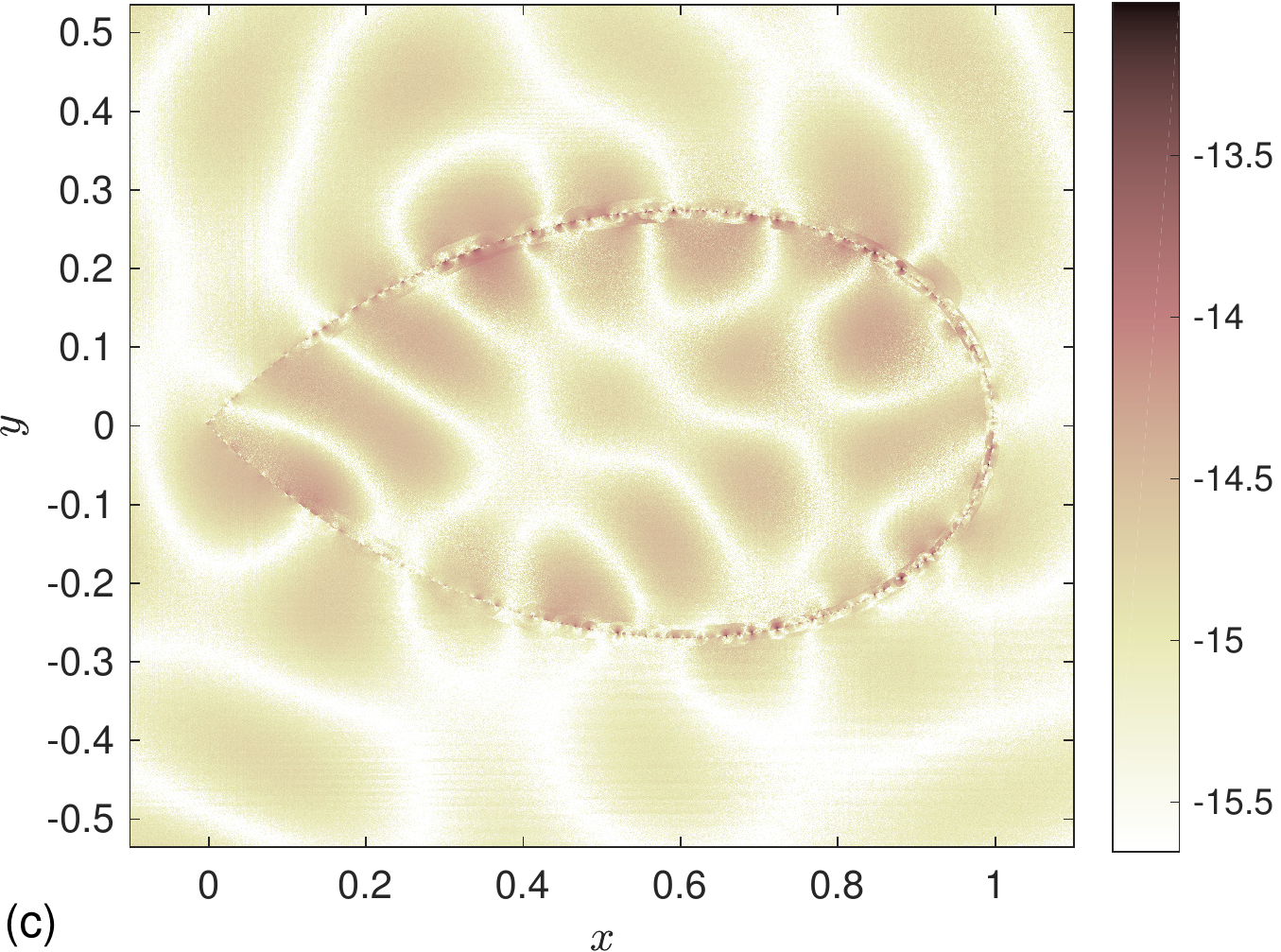}
  \includegraphics[height=47mm]{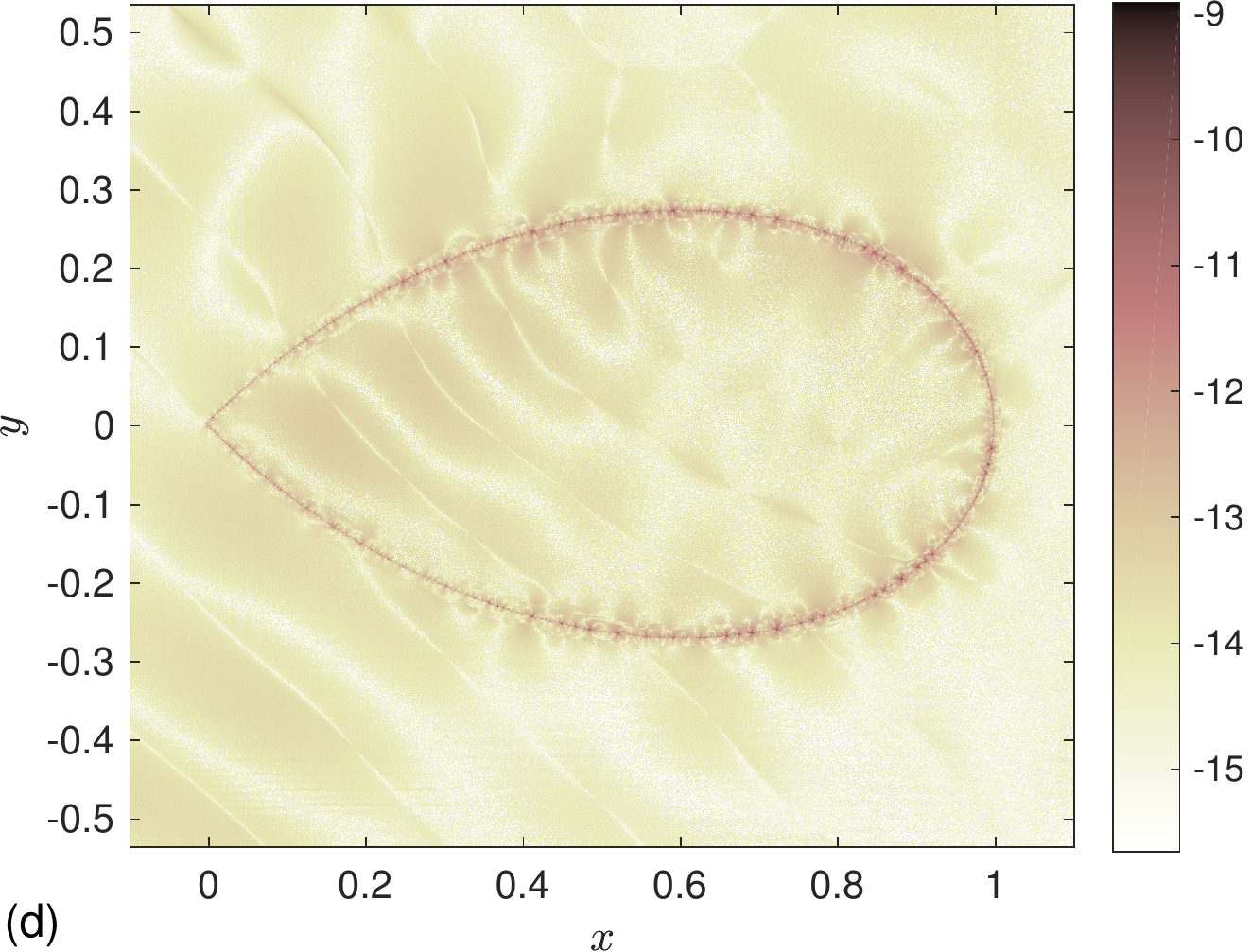}
  \includegraphics[height=46.5mm]{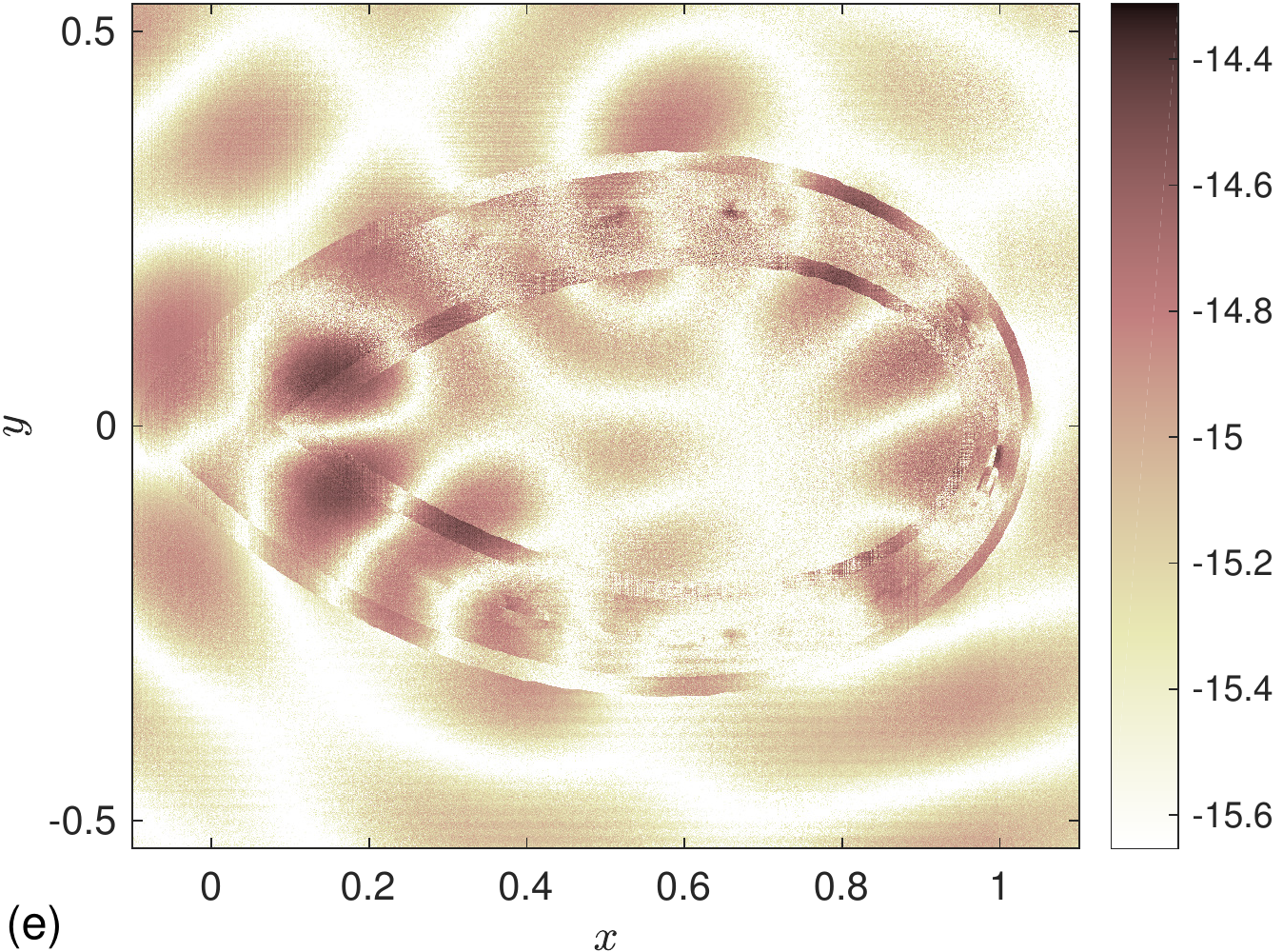}
  \includegraphics[height=46.5mm]{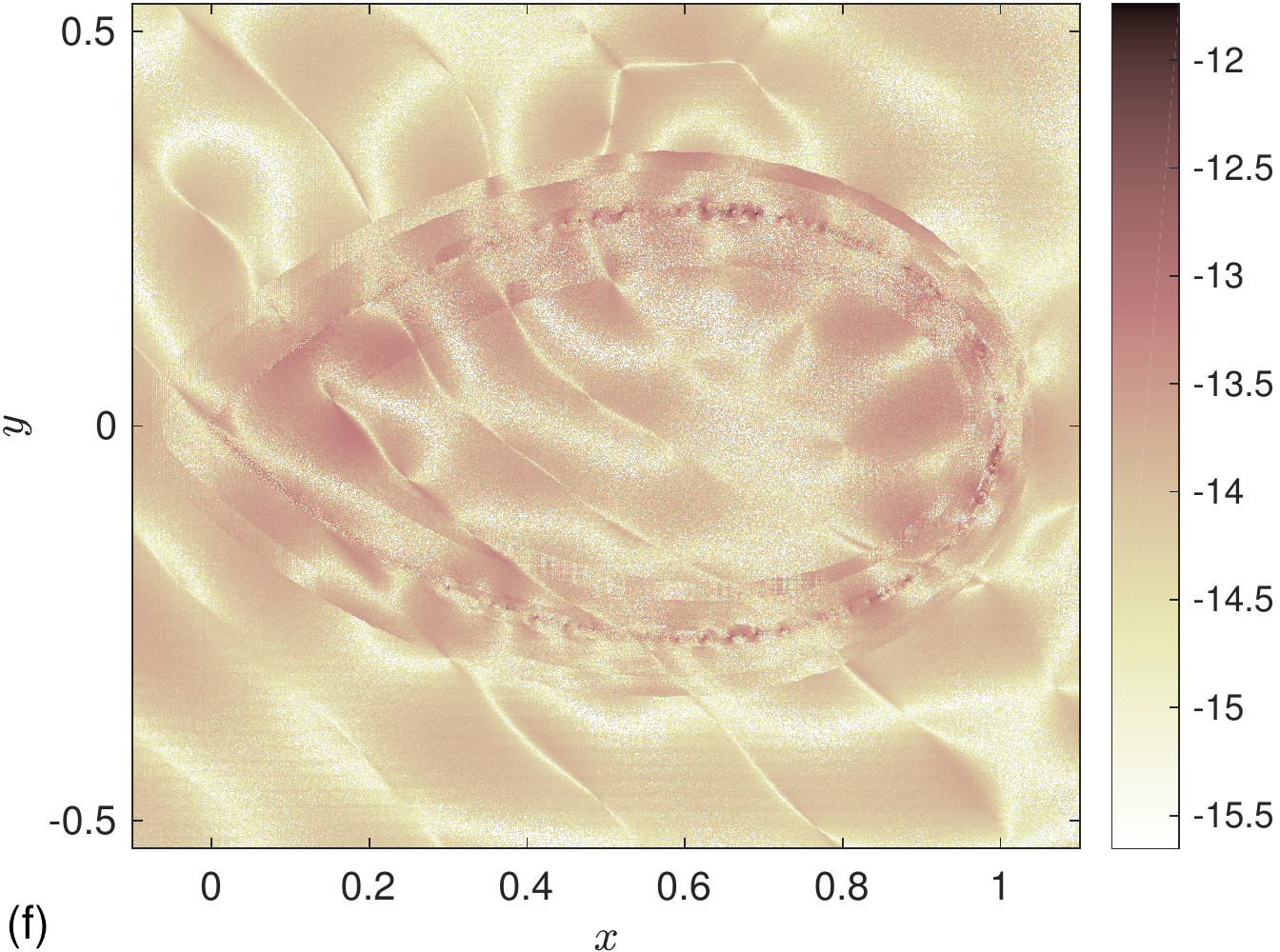}
\caption{\sf $H(r,0)$ and $\nabla H(r,0)$ with $k_1=18$,
  $\varepsilon=2.25$, $\theta=\pi/2$, and
  $d=\left(\cos(\pi/4),\sin(\pi/4)\right)$; (a) The field $H(r,0)$;
  (b) The field $\lvert\nabla H(r,0)\rvert$; (c) $\log_{10}$ of
  estimated absolute error in $H(r,0)$ with KM1; (d) $\log_{10}$ of
  estimated absolute error in $\lvert\nabla H(r,0)\rvert$ with KM1;
  (e) $\log_{10}$ of estimated absolute error in $H(r,0)$ with KM2
  and~(\ref{eq:U3KM2}); (f) $\log_{10}$ of estimated absolute error in
  $\lvert\nabla H(r,0)\rvert$ with KM2 and~(\ref{eq:U3KM2}).}
\label{fig:pos}
\end{figure}

\subsection{Fields and gradient fields}
\label{sec:fgf}

We compare the performance of the KM1 equations to the performance of
the KM2 equations where~(\ref{eq:U3KM2}) is used for field evaluations
at points $r$ close to $\Gamma$. 

The first setup has $\varepsilon=2.25$, $k_1=18$, $\Gamma$ as
in~(\ref{eq:Gamma}), $\theta=\pi/2$, $U^{\rm in}$ as
in~(\ref{eq:Uin}), and $d=\left(\cos(\pi/4),\sin(\pi/4)\right)$.
(Recall that with both $\varepsilon$ and $k_1$ real and positive,
$c=1$ in~(\ref{eq:cchoice}) and the KM1 equations coincide with the
KM0 equations). Both the field $H(r,0)$ and the gradient field $\nabla
H(r,0)$ are computed. Figure~\ref{fig:pos} shows that the achievable
accuracy in $H(r,0)$ and $\nabla H(r,0)$ is improved with around one
and three digits, respectively, when the KM2 system
with~(\ref{eq:U3KM2}) is used rather than the KM1 equations. There are
800 discretization points on the coarse mesh on $\Gamma$ and $10^6$
field points on a (rectangular) Cartesian grid in the box ${\cal
  B}=\left\{ -0.1\le x\le 1.1, -0.54\le y\le 0.54\right\}$.

\begin{figure}[t]
\centering 
  \includegraphics[height=48mm]{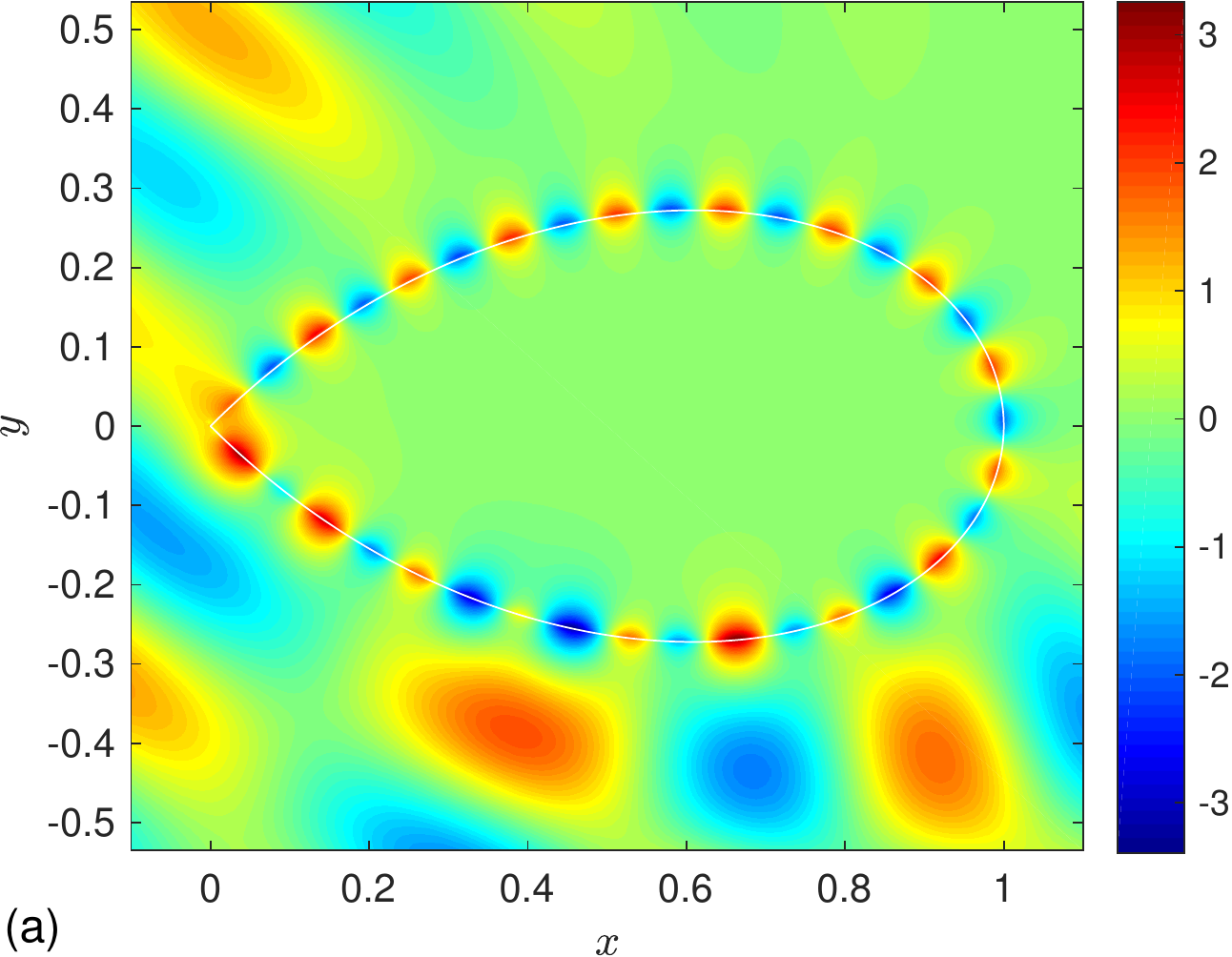}
  \includegraphics[height=48mm]{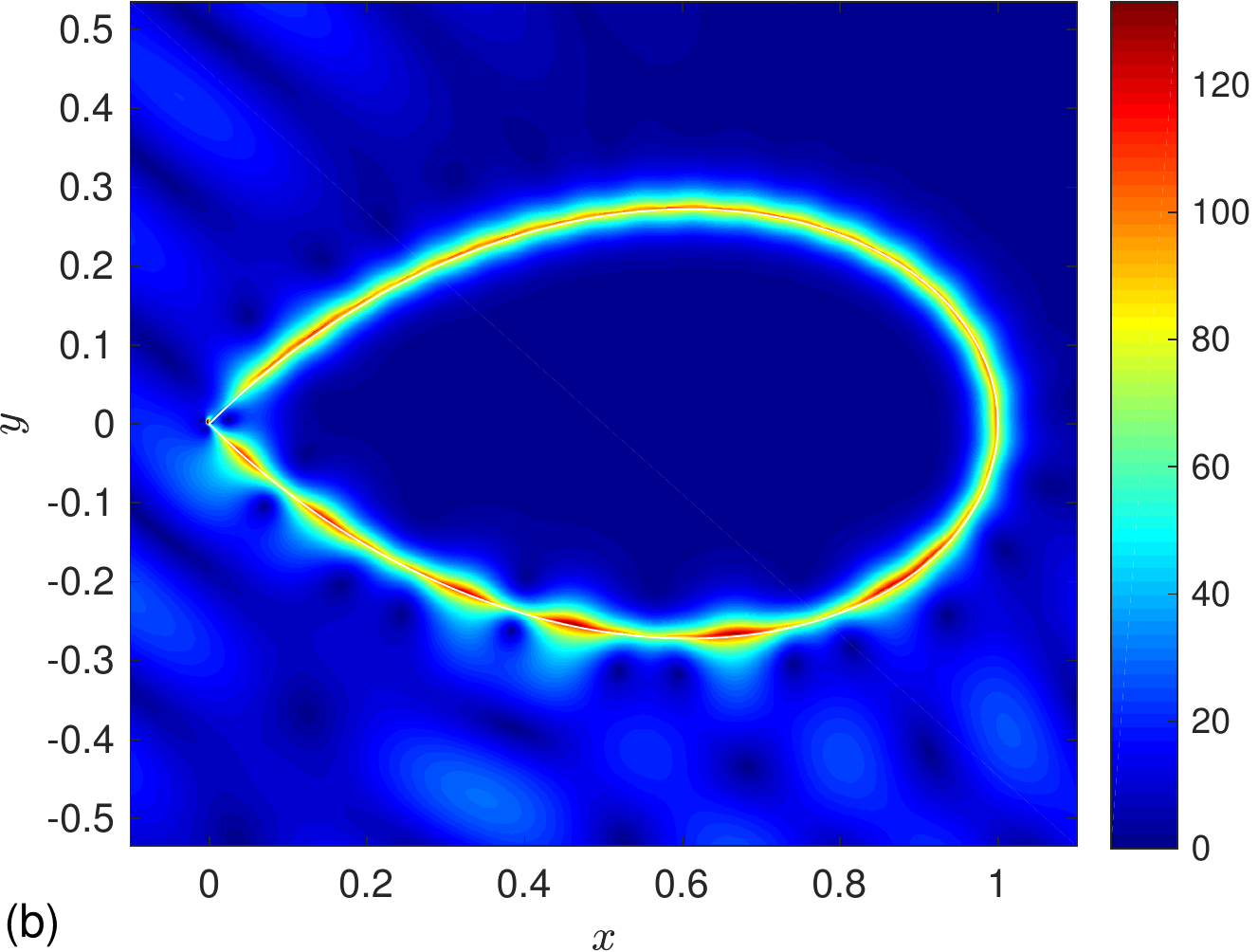}
  \includegraphics[height=47mm]{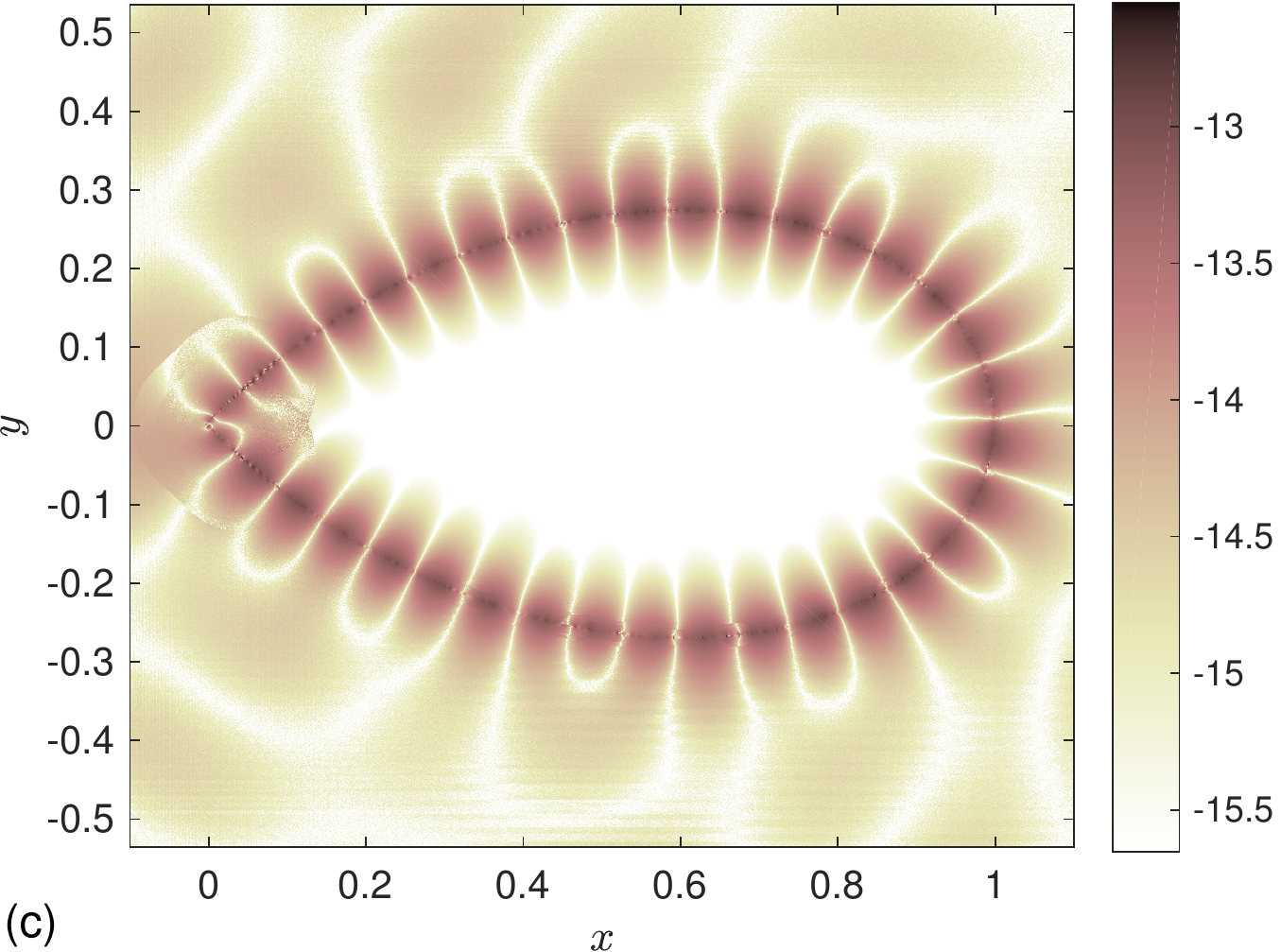}
  \includegraphics[height=47mm]{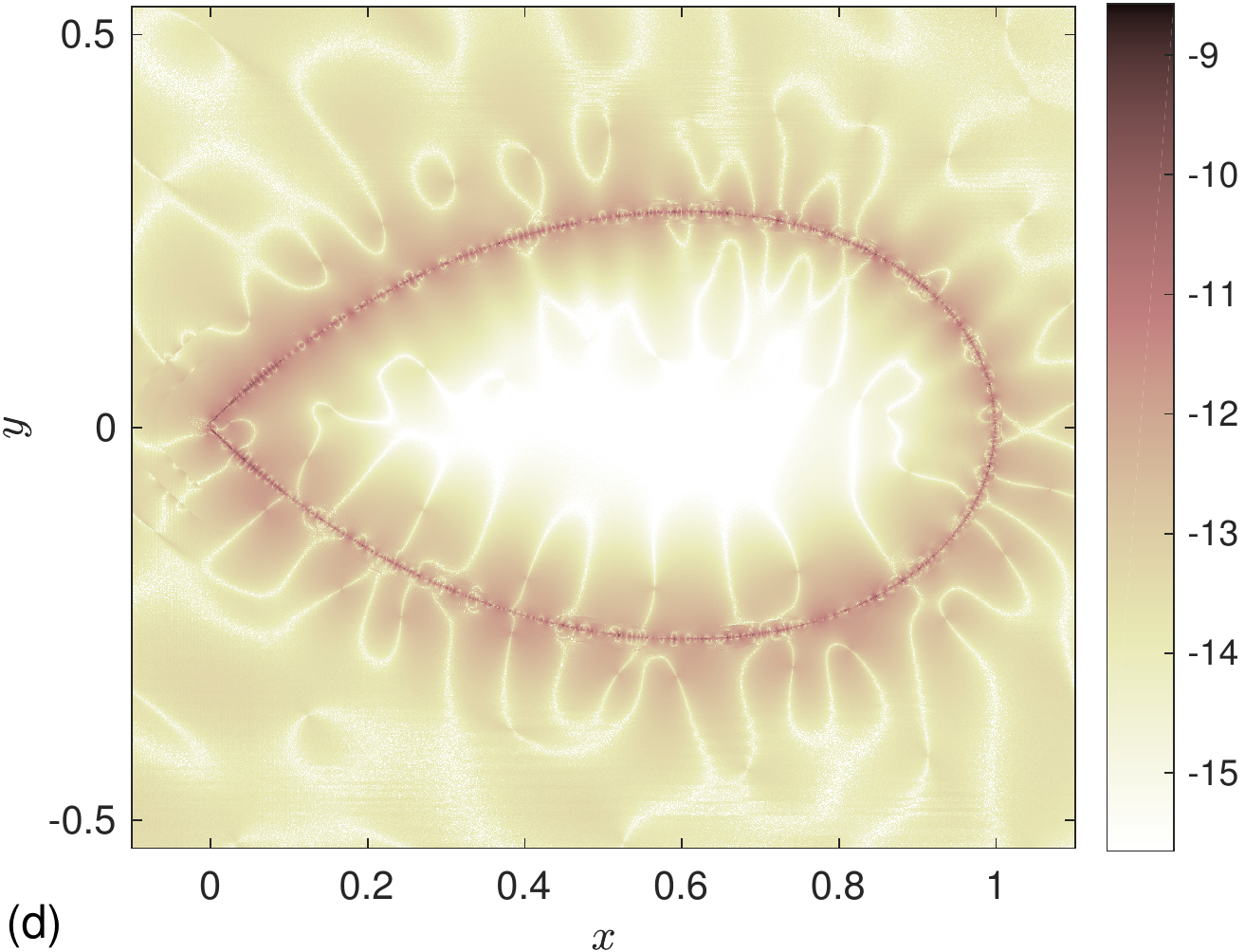}
\caption{\sf $H^+(r,0)$ and $\nabla H^+(r,0)$ with $k_1=18$,
  $\varepsilon=-1.1838$, $\theta=\pi/2$, and
  $d=\left(\cos(\pi/4),\sin(\pi/4)\right)$: (a) The field $H^+(r,0)$;
  (b) The (diverging) field $\lvert\nabla H^+(r,0)\rvert$ with
  colorbar range set to $[0,133]$; (c) $\log_{10}$ of estimated
  absolute error in $H^+(r,0)$ with KM1; (d) $\log_{10}$ of estimated
  absolute error in $\lvert\nabla H^+(r,0)\rvert$ with KM1.}
\label{fig:neg}
\end{figure}

Figure~\ref{fig:neg} shows results for a second setup with
$\varepsilon=-1.1838$, the geometry and the mesh being the same as
above. We only use the KM1 equations, with $c=-{\rm i}$ according
to~(\ref{eq:cchoice}). The surface plasmon wavelength of $2\pi/k_{\rm
  sp}\approx 0.138$ corresponds to $17.6$ wavelengths along $\Gamma$
-- a number that agrees well with the wave pattern of
Figure~\ref{fig:neg}(a). Note that the gradient field $\nabla
H^+(r,0)$ of Figure~\ref{fig:neg}(b) diverges in the corner and that
the colorbar range is limited to $[0,133]$ as to provide full dynamic
color range away from the corner. The absolute gradient field error
shown in Figure~\ref{fig:neg}(d) corresponds to ten digit accuracy or
better.

We end this section with some timings for the computations used to
produce Figure~\ref{fig:neg}(a): setting up the discretized KM1 system
matrix took $4.3$ seconds; constructing the compressed weighted
inverse used for RCIP acceleration took $23$ seconds; solving the main
linear system took $0.25$ seconds; evaluating $H^+(r,0)$ took, on
average, $0.6$ milliseconds per field point in ${\cal B}$.

\begin{figure}[t]
\centering 
  \includegraphics[height=49mm]{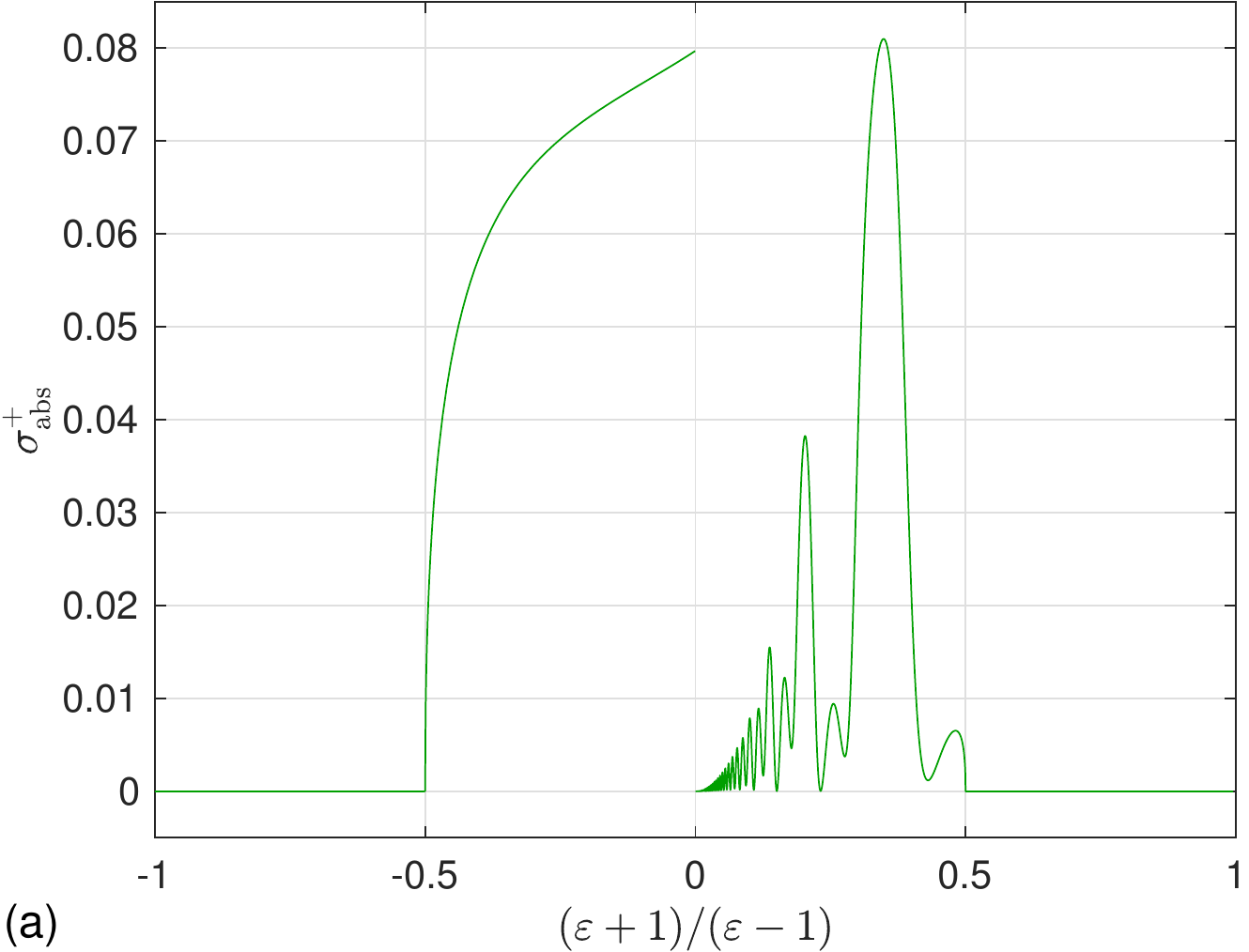}
  \includegraphics[height=49mm]{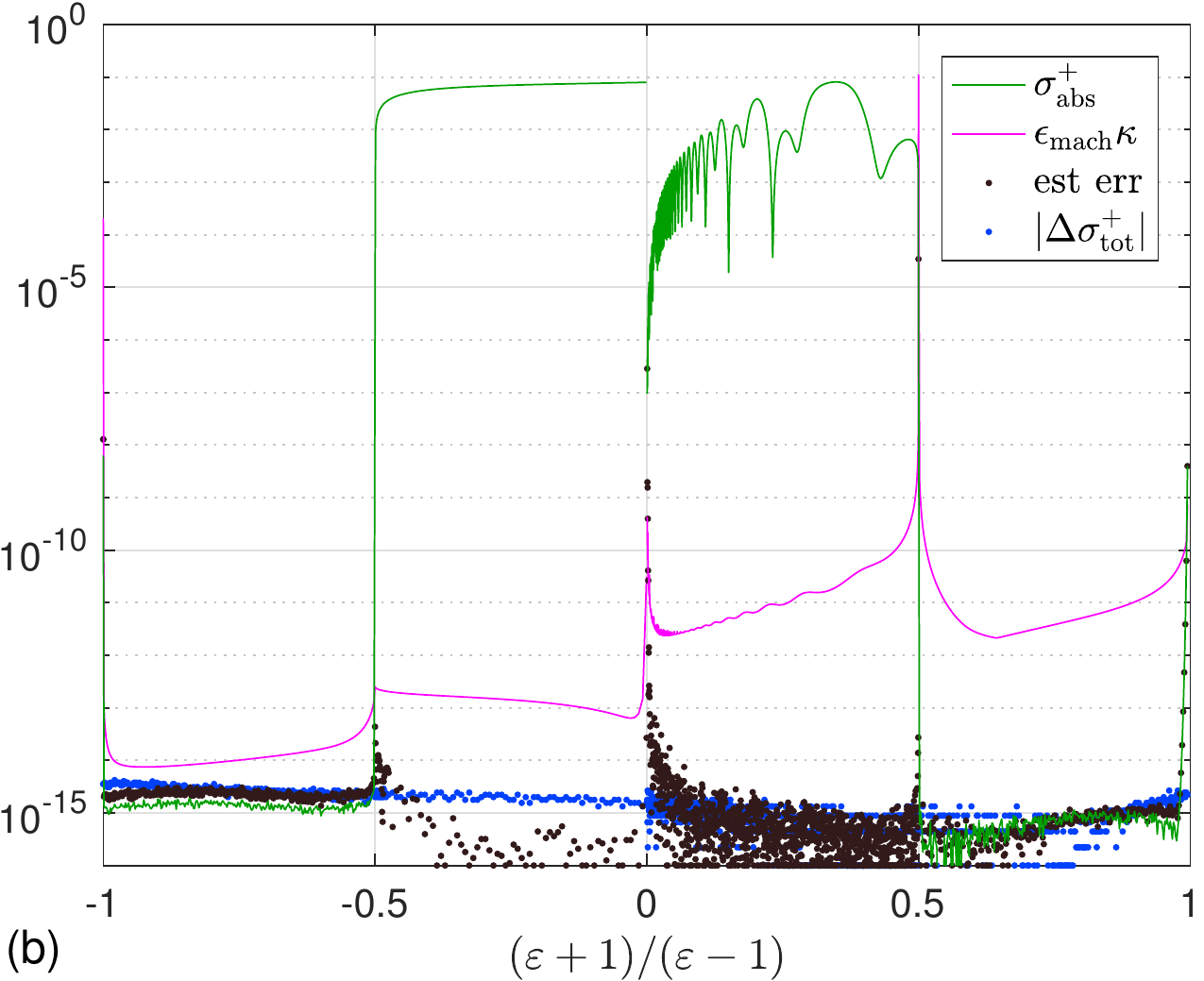}
  \includegraphics[height=51mm]{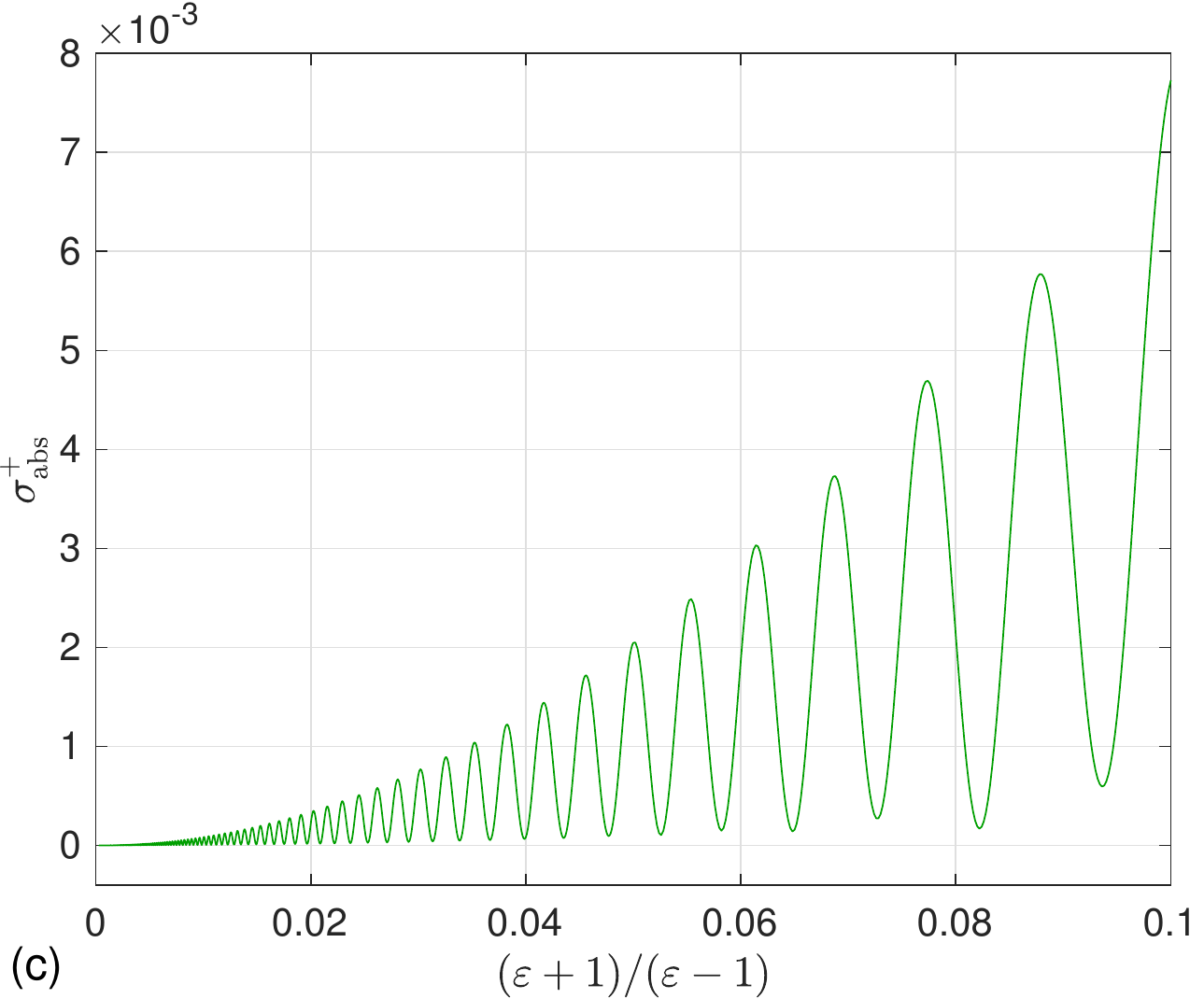}
  \includegraphics[height=49mm]{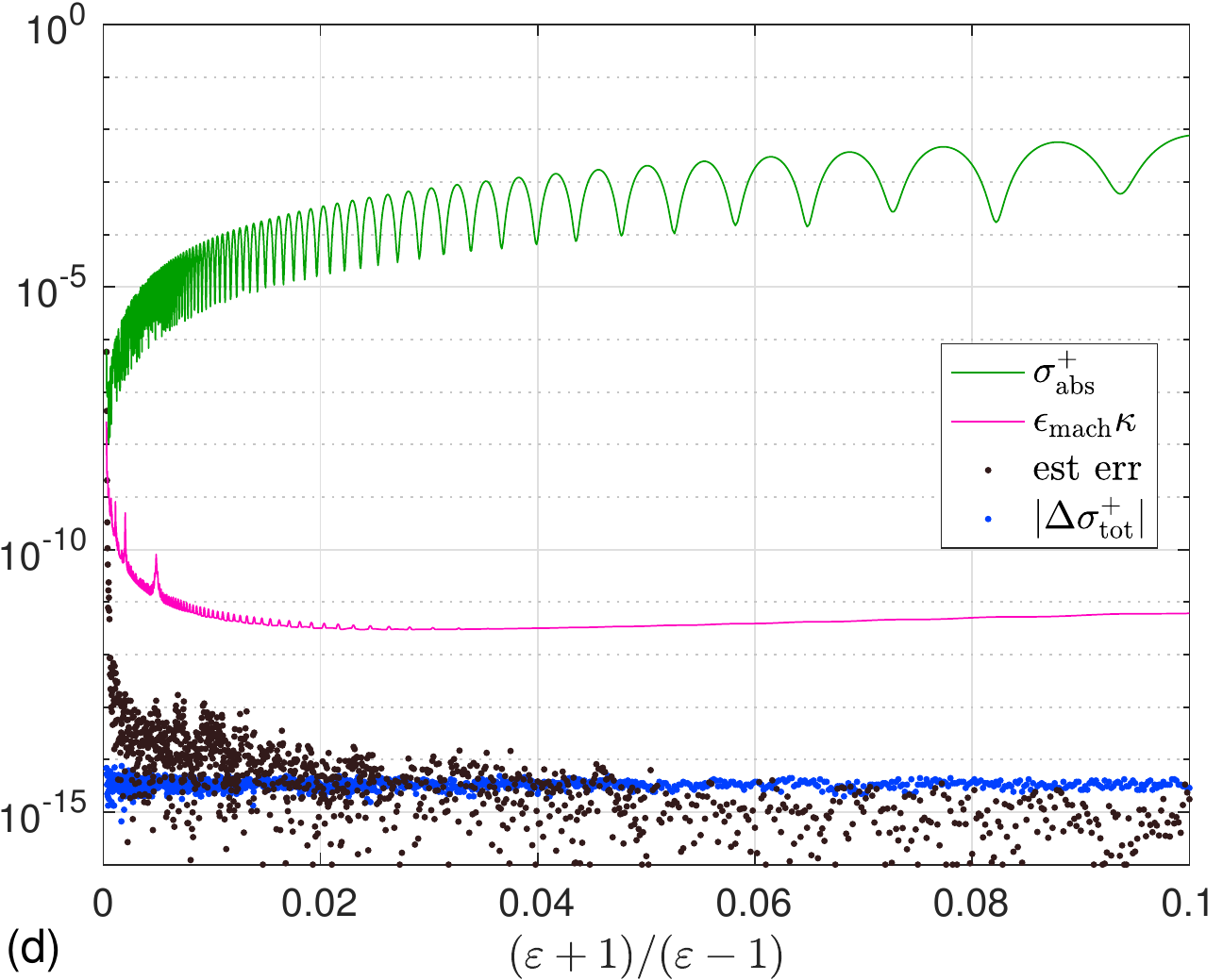}
\caption{\sf The absorption cross section $\sigma_{\rm abs}^+$ 
  of~(\ref{eq:abs}) of an object with $\Gamma$ as in~(\ref{eq:Gamma}),
  $\theta=\pi/2$, $d=\left(\cos(\pi/4),\sin(\pi/4)\right)$, $k_1=18$,
  and $\varepsilon<0$: (a) $\sigma_{\rm abs}^+$; (b) $\sigma_{\rm
    abs}^+$, $\epsilon_{\rm mach}$ times the condition number
  $\kappa$, estimated absolute error in $\sigma_{\rm abs}^+$, and the
  absolute difference of $\sigma_{\rm tot}^+$ from~(\ref{eq:stot1})
  and from~(\ref{eq:stot2}) with logarithmic scale on the $y$-axis;
  (c,d) higher resolution of (a,b) in the interval
  $0<(\varepsilon+1)/(\varepsilon-1)\le 0.1$.}
\label{fig:abs}
\end{figure}

\subsection{The absorption cross section}
\label{sec:absorp}

We compute the limit absorption cross section $\sigma_{\rm abs}^+$
of~(\ref{eq:abs}) for a scatterer with $\Gamma$ as
in~(\ref{eq:Gamma}), $\theta=\pi/2$,
$d=\left(\cos(\pi/4),\sin(\pi/4)\right)$, $k_1=18$, $\varepsilon<0$,
and $\Gamma_{\rm C}$ as the unit circle centered at $r=0.5$. The KM1
equations are used, with $c=-{\rm i}$ according to~(\ref{eq:cchoice}).
Figure~\ref{fig:abs} shows results for $\sigma_{\rm abs}^+$, for a
quantity that is machine epsilon $\epsilon_{\rm mach}$ times the
condition number $\kappa$ of the system matrix, for the estimated
absolute error in $\sigma_{\rm abs}^+$ and, as a consistency check,
for the absolute difference between $\sigma_{\rm tot}^+$ computed
from~(\ref{eq:stot1}) and from~(\ref{eq:stot2}).

The curves in Figure~\ref{fig:abs}(a,b) are resolved by $1839$
different values of $\varepsilon$, $1280$ discretization points on the
coarse mesh on $\Gamma$, and with $n_{\rm pt}=16$ in the underlying
quadrature. The magnified curves in Figure~\ref{fig:abs}(c,d) use
$1239$ different values of $\varepsilon$, $2560$ discretization points
on the coarse mesh, and $n_{\rm pt}=32$. One can see that the
condition number of the discretized KM1 system is low for most values
of $\varepsilon$, with the exception of values that make
$(\varepsilon+1)/(\varepsilon-1)$ belong to the set $\{-1,0,0.5\}$.

Figure~\ref{fig:abs}(d) also shows that high accuracy in $\sigma_{\rm
  abs}^+$ requires at least ten discretization points per surface
plasmon wavelength and that it is not enough to merely resolve the
incident plane wave. For example, with
$(\varepsilon+1)/(\varepsilon-1)=6\cdot 10^{-4}$, which corresponds to
about 200 surface plasmon wavelengths along $\Gamma$ and $12.8$
discretization points per surface plasmon wavelength on the coarse
mesh, the estimated absolute error in $\sigma_{\rm abs}^+$ is around
$10^{-12}$. Smaller values of $(\varepsilon+1)/(\varepsilon-1)$ give a
much larger error since the number of discretization points in
Figure~\ref{fig:abs}(c,d) is fixed while the surface plasmon
wavelength $2\pi/k_{\rm sp}$ decreases with
$(\varepsilon+1)/(\varepsilon-1)$.

The rapid variations in $\sigma_{\rm abs}^+$ for
$0<(\varepsilon+1)/(\varepsilon-1)<0.5$, seen in Figure~\ref{fig:abs},
are due to the coupling between corner fields and surface plasmon
waves. For $-0.5<(\varepsilon+1)/(\varepsilon-1)<0$ there are no
surface plasmon waves and $\sigma_{\rm abs}^+$ varies less. The
general behavior of $\sigma_{\rm abs}^+$ in Figure~\ref{fig:abs} can
be explained using an analytic expression for the absorbed power of
the wedge eigenfields, obtained by inserting~(\ref{eq:Uwedge})
into~(\ref{eq:loss}), multiplied with the squared amplitudes of the
numerically determined corner fields.

\section{Conclusions}

The mathematical and physical theory behind the excitation of surface
plasmon waves in finite metallic objects with sharp edges by incident
plane waves is rather involved. The present work demonstrates that a
robust integral equation-based solver for the underlying Helmholtz
transmission problem can be constructed and used for the detailed
study of surface plasmon waves in difficult situations. The solver
combines a system of integral equations due to Kleinman and Martin,
called KM1 in the present work, with mildly modified off-the-shelf
numerical tools such as Nyström discretization, RCIP acceleration, and
a product integration scheme for the evaluation of layer potentials.
No assumptions about the solution are needed beyond those that are
explicit in the PDE formulation of the problem. Corner fields and
surface plasmon waves can be computed very accurately and that is
crucial for the evaluation and understanding of rapidly varying
absorption cross sections.

The KM1 system contains a parameter $c$, which should be chosen in
agreement with~(\ref{eq:cchoice0}). A choice of $c$ in agreement
with~(\ref{eq:cchoice0}) makes the KM1 system uniquely solvable on
smooth boundaries $\Gamma$ under plasmonic conditions (the incident
wavenumber $k_1$ is real and positive and the permittivity ratio
$\varepsilon$ is real and negative) and when the underlying
transmission problem has a unique solution.  Furthermore, when
$\Gamma$ has corners and $\varepsilon$ is close to (but not on) an
interval on the negative real axis where solutions do not exist, this
choice of $c$ makes one of the layer densities of the KM1 system
(denoted $\rho$ in the present work) particularly easy to resolve
numerically.

As a ``take-home message'' one can say that it is important to choose
$c$ in agreement with~(\ref{eq:cchoice0}) for the KM1 system. The
choice $c=1$, which is common in the literature and gives the KM0
equations of the present work, is only guaranteed to be good when
$k_1$ and $\varepsilon$ both are real and positive. On the other hand,
under such conditions the equations called KM2 in the present work are
preferable. The KM2 and KM0 systems have similar spectral properties,
but the KM2 representation of the total magnetic field $U$ lends
itself better to accurate field evaluation close to $\Gamma$ than does
the representation of $U$ in the KM0 equations.

\section*{Acknowledgement}

\noindent
This work was supported by the Swedish Research Council under contract
621-2014-5159.


\begin{small}
\bibliography{helsbib}
\bibliographystyle{plain}
\end{small}

\end{document}